\documentclass[aps, prb, reprint,showpacs,superscriptaddress,
footinbib,citeautoscript]{revtex4-2}

\usepackage{dcolumn}
\usepackage{color,soul}
\usepackage{graphicx}
\usepackage[dvipsnames]{xcolor}
\usepackage{hyperref}
\usepackage{multirow}
\usepackage{braket}
\usepackage{amsmath}
\usepackage{bm}
\usepackage{amssymb}
\usepackage[titletoc,title]{appendix}
\hypersetup{
	colorlinks=true, 
	citecolor=blue, 
	urlcolor=blue, 
	linkcolor=red
}

\usepackage{soul} %\hl{#1}
\usepackage{xcolor}

\begin{document}
\title{Nonlinear chiral photocurrent in parity-violating magnetic Weyl semimetals}
\author{Shiva Heidari}
\email{shiva.heidari@ipm.ir}
\affiliation{School of Physics, Institute for Research in Fundamental Sciences, IPM, Tehran, 19395-5531, Iran}
\author{Reza Asgari}
\email{r.asgari@unsw.edu.au, asgari@ipm.ir}
\affiliation{School of Physics, Institute for Research in Fundamental Sciences, IPM, Tehran, 19395-5531, Iran}
\affiliation{School  of  Physics,  University  of  New  South  Wales,  Kensington,  NSW  2052,  Australia}
\affiliation{ARC Centre of Excellence in Future Low-Energy Electronics Technologies,  UNSW Node,  Sydney 2052,  Australia}
\begin{abstract}
	The strong correlation between the non-trivial band topology and the magnetic texture makes magnetic Weyl semimetals excellent candidates for the manipulation and detection of magnetization dynamics. The parity violation together with the Pauli blocking cause only one Weyl node to contribute to the photocurrent response, which in turn affects the magnetic texture due to the \textit{spin transfer torque}. Utilizing the Landau-Lifshitz-Gilbert equation and the spin-transfer torque in non-centrosymmetric Weyl magnets, we show that the chiral photocurrent rotates the magnetization from the easy $c$ axis to the $a$ or $b$ axis, which leads to an exotic current next to the photocurrent response. The chiral photocurrent is calculated in the context of quantum kinetic theory and it has a strong resonance on the order of mA/W near the Weyl nodes, the magnitude of which is controlled by the momentum relaxation time. Remarkably, we study the influence of magnetic texture dynamics on the topological nonlinear photocurrent response, including shift and injection currents along with the new chiral photocurrent, and show that both the magnitude and the in-plane orientation of the chiral photocurrent are strongly correlated with the direction of the magnetic moments.
\end{abstract}
\maketitle
% the last sentence suggests that there is a way to distinguish the chiral current from the injection current, am i right?
%=====================
\section{Introduction}
\label{sec:intro} 
The magnetic topological materials provide an ideal platform for rich and fundamental scientific discoveries stemming from the interplay of topology and magnetism \cite{Tokura_2019,Wang_2021,Fan_2016}. An important advance is the realization of magnetic Weyl semimetals \cite{Liu_2019,Morali_2019,Sanchez_2020}, which are distinguished among topological materials by the dynamic interplay of magnetic texture and topological band crossings \cite{Howlader_2020,Lee_2022,Destraz_2020}. The topological notion of Weyl nodes can be understood as effective magnetic monopoles with opposite charges in momentum space as the origin of the diverging Berry curvature \cite{Manna_2018}, leading to the large intrinsic anomalous Hall conductivity \cite{Meng_2019,K_bler_2014,Liu_2018}. The other quantum metrics as the geometric origin of the nonlinear photocurrent response appear in inversion-asymmetric topological materials~\cite{PhysRevX.10.041041,nagaosa2020transport}. The parity-violation in Weyl systems leads to an inverted asymmetric transition of the electron position and velocity in the nonlinear optical response, resulting in the shift and injection currents \cite{PhysRevX.10.041041,PhysRevX.11.011001}. These nonlinear photocurrent conductivities are thoroughly governed by system symmetries that dictate the divergent behavior of topological quantum geometries near the Weyl crossing points. The recently discovered parity-violating magnetic Weyl semimetals are proposed as promising candidates to generate both shift and injection currents due to parity-time (PT) symmetry breaking. In TaAs Weyl semimetals the non-liner chiral photocurrent can be induced by a femtosecond circularly polarized (CP) pulse through the non-equilibrium chiral magnetic effect \cite{PhysRevB.93.201202}.

The magnetic texture in magnetic materials can be controlled by a spin-polarized current without a magnetic field \cite{PhysRevB.54.9353,PhysRevLett.80.4281,PhysRevLett.92.027201,Kiselev_2003,Krivorotov_2005}. The tunable magnetization at the interface between a topological insulator and a ferromagnetic insulator opens up an intriguing venue to discover the intimate relation between non-trivial band topology and magnetic configuration \cite{PhysRevLett.104.146802,PhysRevB.81.121401,PhysRevLett.109.237203,PhysRevB.82.161401,PhysRevLett.108.187201,PhysRevB.89.024413,PhysRevB.90.041412}. The significant role of strong spin-orbit coupling in magnetic Weyl semimetals has attracted much attention in more efficient spintronic applications and magnetic dynamics detection \cite{PhysRevB.104.235119,Suzuki_2019,Howlader_2020}. In particular, the electrically induced structural phase transition in domain walls in a magnetic Weyl semimetal is accompanied by transient nonlinear electrical signals $J_y \propto (\nabla \times \hat{M}(x,t)) E_x^2$, which can be taken as evidence for the magnetic dynamics \cite{heidari2022probing}. Further, the nonlinear optical response in non-centrosymmetric topological magnets can be employed as a powerful tool for realizing the intrinsic connection among the optical response, quantum geometry in momentum space and magnetic texture in real space. Specifically, in a Weyl magnet, the magnitude and direction of magnetization determine the spacing and orientation of a pair of Weyl nodes of opposite chirality in momentum space. Therefore, the potential to control the magnetization can cause a topological redistribution of Weyl nodes in k-space, leading to a remarkable change in the nonlinear shift and injection currents.

In this research, we investigate the optical manipulation of the magnetic texture in an inversion-asymmetric Weyl magnet. The strong spin-orbit coupling in such materials induces a spin-transfer torque (STT) arising from the optical transitions of a single Weyl node. Our key question here is how a luminous light can rotate the magnetization and how these magnetic dynamics then lead to the significant changes in the nonlinear photocurrent responses. The Weyl nodes in an inversion-asymmetric Weyl semimetal do not have the same energy, then the light-induced resonant transitions in one node are Pauli-blocked in another node. Therefore, in the frequency window facing a node, only specific Weyl fermions with chirality $\chi$ contribute to the interband transitions and correspondingly generate a chiral photocurrent response $\bm J_5$. Such a chiral current can interact with the magnetic texture and lead to magnetic rotation. We show that the direction of magnetization rotates from the initial $c$ axis to the final $a$ or $b$ axis, leading to significant changes in magnitude and orientation of the in-plane photocurrent responses.
Our research provides new insights into understanding the role of the tunable magnetic direction in constructing the nonlinear topological shift and injection currents.

The paper is structured as follows. In Sec. \ref{sec2} we introduce the features and symmetries of parity-violating magnetic Weyl semimetals and provide some examples of theoretically predicted and experimentally observed such materials. Sec. \ref{sec3} predicts the light-induced magnetization rotation leading to an additional current owing to the strong coupling between topological Weyl fermions and magnetic orientation. The general properties of the chiral photocurrent response are discussed in Sec. \ref{sec4} and Sec. \ref{sec5} is devoted to the quantum kinetic theory of the second-order DC photocurrent, which stems from the non-trivial quantum geometry of the band structure. The nonlinear Drude response arises from the intraband transitions, and the interband shift, gyration and injection currents and the influence of magnetic dynamics on them are discussed in Sec. \ref{sec6} and Sec. \ref{sec7}. We conclude our findings in Sect. \ref{sec8}.
%...........
\section{Parity-violating Magnetic Weyl semimetals} \label{sec2}
Weyl semimetals have been proposed and have emerged in band structures in which either time-reversal (TR) or inversion (I) symmetries breaks. It has been theoretically proposed and experimentally confirmed that a large class of Weyl materials in RAlX (R=rare earths, Al, X=Si, Ge) realize the Weyl fermions, respecting or violating TR symmetry or I symmetry depending on the choice of rare earth components and can even be categorized into type I or type II Weyl semimetals \cite{Sanchez_2020,Destraz_2020,PhysRevB.97.041104}. More specifically, (Pr,Ce)AlX can be ferromagnetic with an easy axis along the $c$ and $a$ directions \cite{BOBEV20052091,FLANDORFER1998191,GLADYSHEVSKII2000265,DHAR199622,Puphal_2020}, while LaAlGe is nonmagnetic \cite {Xu_2017}.
%===============
%\begin{widetext}
\begin{table}[htbp]
	\caption{Some distinguished examples of experimentally observed or theoretically proposed magnetic and non-magnetic Weyl semimetals. The last two compounds denoted by $^*$, RAlX (R=(Pr, Ce), X=(Ge,Si)), represent three examples of theoretically proposed and experimentally observed parity-violating magnetic Weyl semimetals. } \label{tab1}  % title of Table 
	\centering
	\resizebox{\columnwidth}{!}	{\begin{tabular}{|c|c|c|c|c|c|c|c|c|}
			%\hline
			%\multicolumn{9}{ |c| }{Outline view of results} \\
			\hline
			\multirow{1}{*}{Compounds} & TRS & IS & Type &  Magnetic Properties    \\
			\hline
			\multirow{1}{*}{AB \cite{PhysRevX.5.031013,PhysRevB.92.115428,PhysRevB.92.235104,PhysRevLett.116.066601}} & $\surd$ & $\times$ &  I & Non-Magnetic  \\
			\multirow{1}{*}{A=(Ta,Nb),B=(As,P)} &  &  &   &   \\
			\hline
			\multirow{1}{*}{(W,Mo)Te$_2$\cite{PhysRevB.94.085127,PhysRevB.94.195134}/Ta$_3$S$_2$ \cite{Chang_2016}} & $\surd$ & $\times$ &  II & Non-Magnetic  \\ 
			\hline
			\multirow{1}{*}{LaAlGe \cite{Xu_2017}} & $\surd$ & $\times$ &  II & Non-Magnetic  \\
			\hline
			\multirow{1}{*}{Mn$_3$A} & $\times$ &$\surd$  & II  & Non-Collinear  \\
			\multirow{1}{*}{A=(Sn \cite{Kuroda_2017},Ge \cite{Nayak_2016})} &  &  &   & Anti-Ferromagnetic  \\
			\hline
			\multirow{1}{*}{CuMnSb} & $\times$ &$\times$  & II  & Collinear  \\
			\multirow{1}{*}{\cite{Endo_1968,Regnat_2018,Manna_2018}} &  &  &   & Anti-Ferromagnetic  \\
			\hline
			\multirow{1}{*}{YbMnBi$_2$} & $\times$ &$\surd$  & II  & Canted  \\
			\multirow{1}{*}{\cite{Borisenko_2019}} &  &  &   & Anti-Ferromagnetic  \\
			\hline
			\multirow{1}{*}{Alternative Layers of Magnetically } & $\times$ & $\times$ &  I & Magnetic Impurities \\
			\multirow{1}{*}{Doped TI and NI as the Spacer \cite{PhysRevLett.107.127205}} &  &  &   & Order Ferromagnetically  \\
			\hline
			\multirow{1}{*}{Co$_3$Sn$_2$S$_2$ \cite{Morali_2019}/HgCr$_2$Se$_4$ \cite{PhysRevLett.107.186806}} & $\times$ & $\surd$ &  I & Ferromagnetic  \\
			\hline
			\multirow{1}{*}{PrAlSi \cite{PhysRevB.102.085143,PhysRevB.104.014412}} & $\times$ & $\surd$ &  I & Ferromagnetic  \\
			\hline
			\multirow{1}{*}{PrAlX$^*$} & $\times$ & $\times$ &  I & Ferromagnetic  \\
			\multirow{1}{*}{X=(Ge \cite{Sanchez_2020,PhysRevB.97.041104},Si \cite{sakhya2022observation})} &  &  &   &   \\
			\hline
			\multirow{1}{*}{CeAlGe$^*$\cite{Puphal_2020,PhysRevB.97.041104}} & $\times$ & $\times$ &  II & Ferromagnetic  \\
			\hline
	\end{tabular}}
\end{table}
%	\end{widetext}

In this study, we focus on the ferromagnetic Weyl semimetals, where the intrinsic magnetic texture breaks the time-reversal symmetry below the Curie temperature T$_c$, and also their peculiar lattice crystal breaks the inversion symmetry (IS). Table. \ref{tab1} has classified some examples of proposed or realized Weyl semimetals according to TR, I and Lorentz symmetries (type I or type II) and their magnetic properties. The ferromagnetism in such materials arises from the ordering of the local moments of the f-electrons. For example, in LaAlGe, a non-magnetic Weyl semimetal, the f-orbital in the electronic configuration of the La atom is empty, citing the density of states (DOS) of the localized f-orbital found only in the conduction band. On the other hand, in (Pr,Ce)AlGe, the f-orbital in the Pr (or Ce) atom contains two (or one) electrons, leading to ferromagnetization in these two materials \cite{PhysRevB.97.041104}.
\begin{figure}[t]
	\centering
	\includegraphics[scale=1.2]{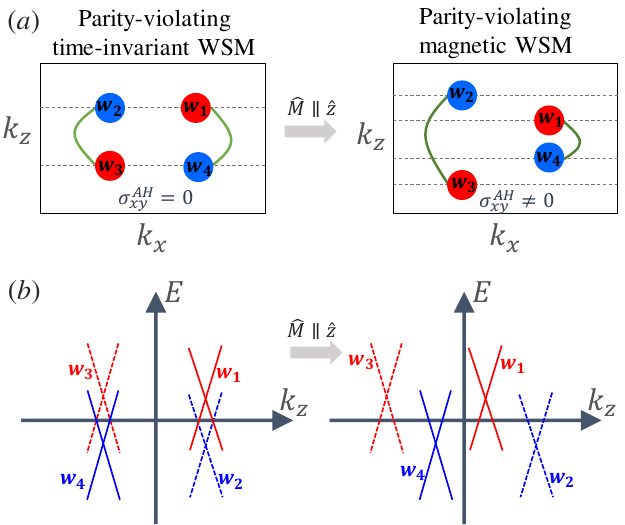} 
	\caption{ (a) Left hand side: The minimum model for a time-invariant inversion asymmetric Weyl semimetal with four nodes. The blue and red circles represent the Weyl nodes with opposite chiralities. Right hand side: The magnetism along $\hat{z}$ rearranges the Weyl nodes to violate TRS. The green lines are the Fermi arcs connecting two Weyl nodes. (b) Represents the corresponding schematics energy dispersion of Weyl nodes in non-magnetic and ferromagnetic Weyl semimetals. The solid lines denoted the Weyl nodes with $k_{x,y}> 0$ and the dashed lines are nodes with $k_{x,y}< 0$. } \label{fig1}
\end{figure}
In magnetic Weyl semimetals with inversion symmetry, i.g.,  Co$_3$Sn$_2$S$_2$ \cite{Morali_2019}, HgCr$_2$Se$_4$ \cite{PhysRevLett.107.186806} and PrAlSi \cite{PhysRevB.102.085143,PhysRevB.104.014412}, the magnetism of texture splits the Weyl nodes in k-space. However, the parity-violating magnetic Weyl semimetals are generated by IS breaking, i.e., nodes are split by inversion symmetry breaking, and the magnetic texture only reconfigures the Weyl nodes along the magnetic direction in order to break TR symmetry.
Figure \ref{fig1} shows how magnetism rearranges the Weyl nodes in an inversion asymmetric crystal.
The low-energy Hamiltonian describing a  minimum model near w$_1$ and w$_4$ nodes is given by [Appendix. \ref{appa}]
\begin{eqnarray}
		H_0 =&\hbar v_{\rm F} ({\bm k}_\perp-\bm{k}_\perp^w) \ \tau_0 \otimes {\bm \sigma}_\perp
		+\hbar v_{\rm F} \ (k_z \tau_z- \\ & (k^w_z-k_M)\tau_0) \otimes \sigma_z +u \ \tau_z \otimes \sigma_0 +\lambda \ \tau_y \otimes \sigma_z. \nonumber
\end{eqnarray}
The last two terms are responsible for inversion symmetry breaking, while other symmetries are preserved. The term $\lambda \tau_y \otimes \sigma_z$ is the momentum-independent spin-orbit interaction that split the degeneracy at every points except the Weyl crossings. This term is closely analogous to the \textit{Dresselhous spin-orbit interaction} term allowed in the absence of IS. The role of $u\tau_z \otimes \sigma_0$ is shifting two tips of Weyl cones in energy and breaks IS.
The positions of the nodes are given by $\mathbf k_{(w_1,w_3)}=(\pm k^w_\perp,\pm k_z^w-k_M)$ and $\mathbf k_{(w_2,w_4)}= (\mp k^w_\perp,\pm k_z^w+k_M)$, where the vector $\mathbf k^w=(k^w_\perp,k_z^w)$ denotes the node positions in time-invariant inversion asymmetric WSM and $\mathbf k_M=({\cal J}S/\hbar v_F) \hat{M}$ is the momentum separation due to magnetism [Fig. \ref{fig1}], where ${\cal J}$ is the ferromagnetic exchange interaction and $S$ is the magnitude of the texture magnetic moment.
In the time-invariant WSMs [Fig. \ref{fig1}: Left], e.g. TaAs family, no linear anomalous Hall effect is observed. On the other hand, parity-violating magnetic Weyl semimetals [Fig. \ref{fig1}: Right], i.e. PrAl(Ge,Si) and CeAlGe, the Chern number is nonzero in the regions between nodes that are displaced by texture magnetic moments. Therefore, the anomalous Hall effect is expected to be present in the linear response to an electric field in parity-violating magnetic WSMs \cite{Meng_2019}. For PrAlGe with node spacing $k_w=0.15$ \AA$^{-3}$ the intrinsic anomalous Hall conductivity is estimated to be $\sigma_{\text{AH}}=738 \ \Omega^{-1} \text{ cm }^{-1}$ \cite{Sanchez_2020}.

The divergent behavior of the quantum geometry of the electron wave function near the Weyl points plays an important role in the bulk photogalvanic effect in topological materials \cite{PhysRevX.10.041041}. The presence of TR or parity-time ({\cal{PT}}) symmetry helps us predict which of the nonlinear response elements may be zero \cite{bhalla2021quantum,PhysRevX.10.041041}. In the presence of the {\cal{PT}} symmetry, the Berry curvature vanishes at any point in $k$ space, so there is no way to get the Weyl phase. MnGeO$_3$ is a 3D Dirac semimetal with both TR and I symmetry broken while {\cal{PT}} symmetry is preserved. However, in parity-violating magnetic Weyl semimetals such as PrAl(Ge,Si) and CeAlGe, both TR and {\cal{PT}} symmetry are broken and both shift and injection currents are expected to be present.  In the following, we will discuss two impactful parameters in the non-linear photocurrent response in Weyl semimetals. The first is the magnetization direction and the second is the chemical potential which determines the energy difference between the Fermi surface and the singularity of quantum geometries.
\section{Chiral current-induced magnetic rotation} \label{sec3}
%The Weyl nodes in an inversion-assymmetric Weyl semimetal are not at the same energy, then the light-induced resonant transitions in one node will be Pauli blocked in another node. Therefore, in the frequency window faces one node, only specific Weyl fermions with chirality $\chi$ contributes to the inter-band transitions, and accordingly generates a chiral photocurrent response $\bm J_5$. Such chiral current can interact with magnetic texture and leads to the magnetic rotation. We will show that the magnetization direction will rotate from the initial $c$-axis to the final $a$- or $b$-axis, resulting in significant changes in the in-plane photocurrent responses.

In this section, we aim to calculate the light-induced \textit{spin-transfer-torque} (STT) $\bm T_e$ through the non-equilibrium spin polarization of electrons $\braket{\sigma (r)}$, which appears when one node contributes more than the other node, leading to the chiral current $\bm J_5=ev_F \braket{\sigma (r)}$. The chiral current can be induced as a non-equilibrium photo-response. Using the well-known \textit{Landau-Lifshitz-Gilbert} (LLG) equation \cite{landau1935theory}
\begin{equation} \label{dyn}
	\dfrac{d \hat{M}}{dt}= \gamma_0 {\bm{ B}}_{\text{eff}}\times {\hat{M}}+\alpha \hat{M}\times\dfrac{d \hat{M}}{dt}+\bm{T}_e,
\end{equation}
the dynamic behavior of the magnetic texture can be extracted. The parameter $\gamma_0$ is the \textit{gyromagnetic ratio}, ${\bf B}_{\text{eff}}$ is an effective magnetic field, $\alpha$ is the \textit{Gilbert} or \textit { viscous damping parameter}, which is proportional to the energy loss rate \cite{malozemoff2016magnetic}, and $\bm{T}_e$ is the STT describing the electronic background contribution to the magnetic texture dynamics.

\begin{figure}[t]
	\centering
	\includegraphics[scale=0.9]{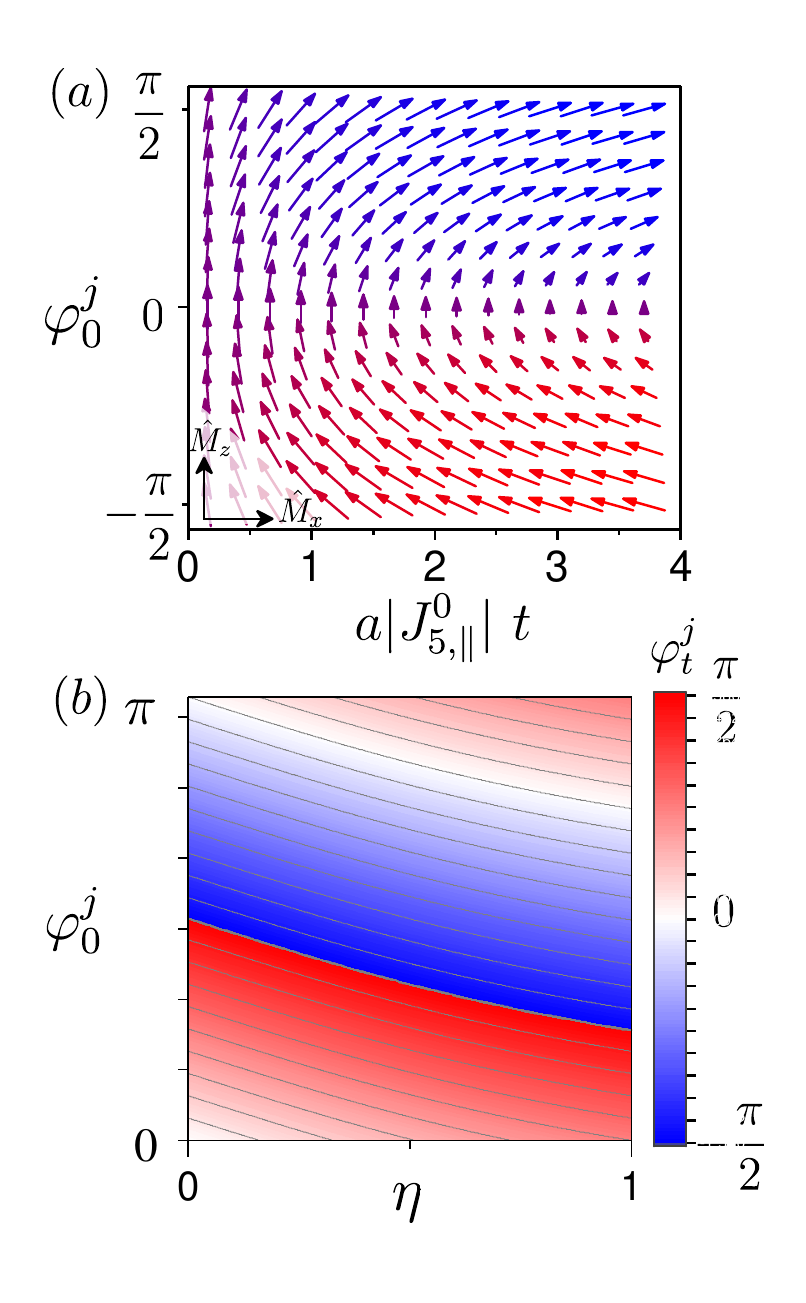} 
	\caption{(a): The light (propagates along the $z$-axis)-induced  magnetic texture evolution in a parity-violating magnetic Weyl semimetals. The magnetization direction rotates away from the initial $c$-axis to the final $a$- or $b$-axis depending on the initial chiral photocurrent magnitude $J^0_{5,\parallel}$ and its orientation $\varphi^j_0$. In the intermediate stage, between the initial and equilibrium states, an axial electric field $\bm E_5$ is induced to the system. We define $a=\frac{{\cal J}S}{e\hbar v_{\rm F} \rho_s}$. (b): The density plot of the modified in-plane photocurrent due to the magnetic texture dynamics. $\varphi_j^0$ and $\varphi_j^t$ are the orientation of in-plane chiral photocurrent in the absence and presence of magnetic dynamics.} \label{fig2}
\end{figure}

\begin{figure}[t]
	\centering
	\includegraphics[scale=0.75]{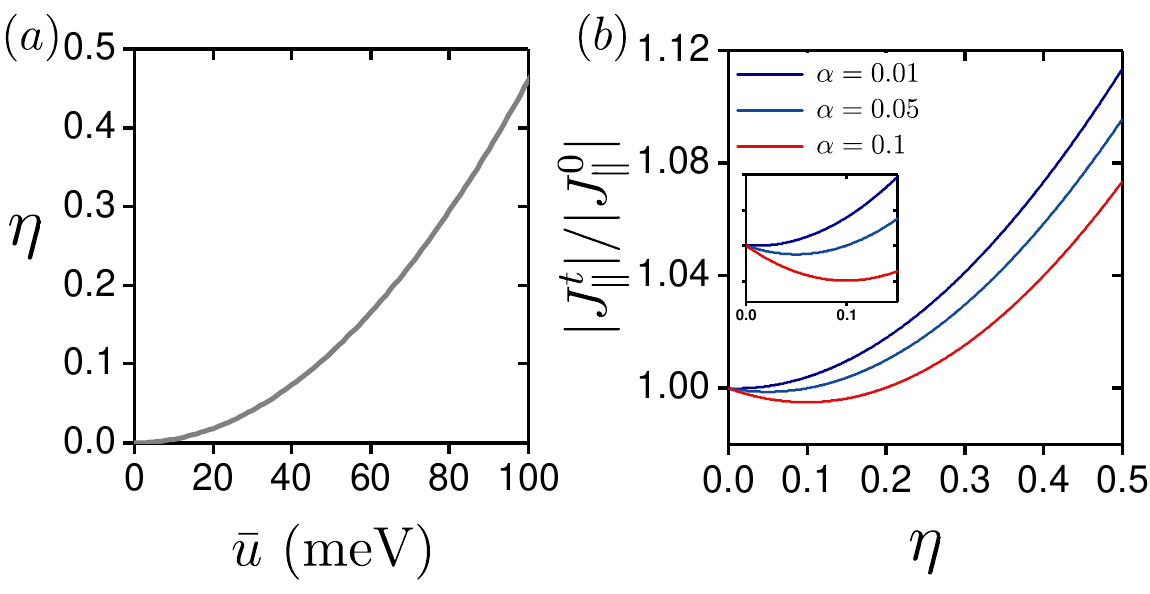} 
	\caption{(a): The parameter $\eta$ with respect to $\bar{u}$ where $\bar{u}=\sqrt{(u_+^2+u_-^2)/2}$. (b): The effect of magnetic texture dynamics on the magnitude of chiral photocurrent. The modification coefficient of chiral current due to magnetic dynamics at time $t$ is given by $|\bm J^t_{5,\parallel}|=\sqrt{(1-\alpha \eta)^2+\eta^2} |\bm J^0_{5,\parallel}|$. The inset demonstrates that in the case of $\alpha>\eta$, the relative magnitude slightly lowers.} \label{fig3}
\end{figure}

The STT $\bm T_e$ on the right-hand side arises from the exchange interaction between itinerant electrons and the magnetic moments and is given by $\bm{T}_e=\frac{|k_M|}{e\rho_s } \hat {M} \times \bm{J}_5,$ where $\bm J_5=\sum\limits_\chi \chi J_\chi$ is the chiral stream. The wave vector $k_M$ determines the separation of the Weyl nodes in momentum space due to magnetism, i.e. $\chi k_M \hat{M}_z$, which is given by $k_M=\frac{{\cal J }S}{\hbar v_F}$, and $\rho_s$ is the number of local magnetic moments per unit volume. Therefore, the STT arises when one node contributes more than the other node in inducing a torque on localized magnetic moments. In other words, the non-equilibrium spin polarization of electrons is essential to rotate the magnetic texture, so the total current $\bm J=\sum\limits_\chi J_\chi$  can not induce STT. Here, the main origin of such a nodal imbalance is the optical transitions of a single Weyl node while the other node is Pauli blocked. We note that because of $\hat{M}(t=0)=\hat{z}$ the STT only applies to the in-plane components of the magnetic moments. Figure \ref{fig2} represents the time evolution of magnetic moments induced by the chiral current from the initial $z$-axis to the final $x$- or $y$-axis, depending on the magnitude of the chiral photocurrent $J^0_{5, \parallel }$ and its orientation $\varphi^j_0$ at $t=0$ [Appendix. \ref{appb}].
Using the LLG equation, and in the absence of a magnetic field $\bm{B}_{\text{eff}}$, we show that strong spin-orbit coupling in magnetic Weyl semimetals together with nonlinear light-induced electronic spin polarization leads to an additional nonlinear photocurrent response generated by magnetic dynamics. In magnetic Weyl semimetals, the magnetic texture in real space is inherent to the nodes position in momentum space and exhibits a dynamical interplay with it \cite{Araki_2019,heidari2022probing}. This key feature leads to an effective U(1) axial vector $\bm{A}_5=({\cal J}S/ev_F)\hat{M}$, in the low energy Hamiltonian [Appendix.\ref{appa}]. Hence, the dynamical behavior of $\bm{A}_5$ gives rise to an axial electric field $\bm E_5=-\frac{{\cal J}S}{e v_F} \dot{\hat{M}}$. Using Eq. (\ref{dyn}), $\bm E_5$ can be written as [Appendix. \ref{appb}] 
\begin{equation} \label{E5}
	\bm E_5=-\dfrac{\hbar |k_M|^2}{e^2\rho_s} \exp(\alpha \hat{\Theta}) (\hat{M}\times \bm J_5),
\end{equation}
where $ \exp(\alpha \hat{\Theta}) {\cal O}=(1+\alpha \hat{M} \times {\cal O}+\alpha^2 \hat{M} \times ( \hat{M} \times {\cal O})+\cdots)$ and $\alpha$ is the dissipation parameter and $\bm J_5$ is the chiral photocurrent response. We should note that the axial electric field $\bm E_5$ only exists in the dynamic regime before approaching the steady state. Such a pseudo-electric field induces a longitudinal drift current $\delta \bm{j}^\chi=\chi \delta \sigma _\chi \bm{E}_5$ for each node $\chi$ \cite{Araki_2019}. The longitudinal conductivity $\delta \sigma_\chi$ is well defined if two Fermi surfaces of two nodes are far enough apart in momentum space that the scattering between the nodes can be neglected. Then the longitudinal conductivity for Weyl semimetals can be estimated as
\begin{equation}
	\delta \sigma_\chi=\dfrac{- e^2 \tau u_\chi^2}{3\pi^2 \hbar^3 v_F}\simeq -(10^4-10^5) \ \text{S/m},
\end{equation}
 where we choose $u_{\pm}=20 \ (\text{or} \ 50)$ meV (the energy separation between nodes and Fermi surfaces), $v_F=5 \times 10^5$ m/s and $\tau=1$ ps. Using the expression for the axial electric field in Eq. (\ref{E5}), we may write the (axial) current $\delta j_{(5)}$ as a consequence of the interplay between electronic degree of freedom and the magnetic texture $\hat{M}$
\begin{equation}
	\begin{split}
		&\delta \bm j= \exp(\alpha \hat{\Theta})  (\hat{M}\times \bm J_5) \sum\limits_\chi \eta_\chi,\\
		&\delta \bm j_{5}= \exp(\alpha \hat{\Theta})  (\hat{M}\times \bm J_5) \sum\limits_\chi \chi \eta_\chi
	\end{split}
\end{equation}
where $\eta_\chi=\dfrac{\tau u_\chi^2 |k_M|^2}{3\pi^2 \hbar^2 v_F \rho_s}$ is a dimensionless quantity and depends on the system-dependent or non-universal parameters.
%	\begin{figure}[t]
	%		\centering
	%		\includegraphics[scale=2.0]{fig7.pdf} 
	%		\caption{The density plot of the modified in-plane photocurrent due to the magnetic-texture dynamics. $\varphi^{e}$ and $\varphi^{e+s}$ are the in-plane orientation of chiral photo-current in the absence and presence of magnetic-texture. } \label{fig7}
	%	\end{figure}
Therefore, the light-induced magnetic dynamics in magnetic Weyl semimetals can induce chiral current $\delta \bm j_5=\delta \bm j_+-\delta \bm j_-$ 
with components
\begin{equation} \label{deltaj}
	\begin{split}
		& \delta j_{5,x}=-\eta (J_{5,y}+\alpha J_{5,x}+{\cal O}(\alpha^2)+\cdots), \\
		& \delta j_{5,y}=\eta (J_{5,x}-\alpha J_{5,y}+{\cal O}(\alpha^2)+\cdots),
	\end{split}
\end{equation}
where $\eta=\eta_++\eta_-$, and $\pm$ stands to opposite chiralities. Figure \ref{fig3}(a) estimates the parameter $\eta$ using the quantities $\tau=1$ ps (the typical relaxation time in semimetals), $v_F=5\times 10^5$ m/s \cite{Behrends_2016}, $k_M=0.15$ \AA$^{-1}$, $\rho_s\simeq1.5 \times 10^{28}$ m$^{-3}$ (volume of unit cell: $V=262.1$ \AA$^{3}$ for a ferromagnetic Weyl semimetal candidate PrAlGe \cite{Meng_2019}). Figure \ref{fig3}(b) represents the change in magnitude of the in-plane chiral current ($x$ and $y$ components) at time $t$, i.e., $\bm{J}_\parallel^t$, in comparison to its magnitude at $t=0$. If $\eta$ exceeds the viscose damping parameter $\alpha$ ($ \eta > \alpha$), the magnetization dynamics increase the magnitude of the in-plane chiral photocurrent response and vice versa in the inset.
Having used Eq. (\ref{deltaj}), we conclude that magnetic texture dynamics can also change the orientation of the in-plane current density as
\begin{equation}
	\tan \varphi_{t}^j=\dfrac{\tan \varphi_0^j+\eta-\alpha \eta \tan \varphi_0^j}{1-\eta \tan \varphi_0^j-\eta \alpha},
\end{equation}
where $\varphi_0^j=\tan^{-1}(J_{5,y}/J_{5,x})$ and $\varphi_{t}^j=\tan^{-1}((J_{5,y}+\delta J_{5,y})/(J_{5,x}+\delta J_{5,x})
)$. Here $\varphi^j_t$ and $\varphi^{j}_0$ are the polar angles between $x$ and $y$ components of current with and without taking into account the magnetic texture dynamics [Fig. \ref{fig2}(b)]. According to Fig. \ref{fig2} (b), the magnetic texture dynamics can reorient and even reverse the in-plane photocurrent. Therefore, the magnetic texture dynamics may lead to a significant change in the orientation of the in-plane chiral photocurrent.

It is worth noting that the above discussion is a general result and can be applied for both linear and circular polarization of light, which can induce chiral current into the system. The spin manipulation in topological materials, particularly Weyl semimetlas, has received considerable attention due to its wide applications in spintronic devices ~\cite{heidari2022probing,Suzuki_2019,Destraz_2020,Hannukainen_2021,Brataas_2012,Lee_2022}.
In the next section, we will discuss the electronic contribution of chiral photocurrent ($J_5^{(2)}$) in the context of the non-linear quantum kinetic theory, and show how the magnitude and direction of the in-plane components can be affected by the direction and magnitude of the magnetic moments.

\section{Chiral photocurrent response} \label{sec4}
Motivated by recent measurements of the nonlinear optical response in transition metal monopnictides such as TaAs, TaP, NbAs and the Weyl semimetal RhSi and CoSi~\cite{wu2017giant, rees2020helicity, ni2021giant}, we perform the chiral photocurrent response in magnetic Weyl semimetals.

The general form of the second-order response to the electric field of light is defined as~\cite{sturman2021photovoltaic,Boyd2007}
\begin{equation}
	\begin{split}
		J^{(2)}_l &(\omega;\omega_1,\omega_2)=\\ & \sum\limits_{i,j} \int \dfrac{d\omega_1 d\omega_2}{(2\pi)^2} \sigma^{l;ij}(\omega;\omega_1,\omega_2) E^i(\omega_1) E^j(\omega_2),
	\end{split}
\end{equation}
where $\omega=\omega_1+\omega_2$. The DC-photocurrent response due to the irradiation of light with frequency $\Omega$ is characterized in the condition of $\omega=0$ or $\omega_1=-\omega_2$. We define $\omega_1=-\omega_2=\Omega$, then we will have $E^i(\Omega) E^j(\omega-\Omega)=E^i(\Omega) E^j(-\Omega)$ or $\omega=0$ in our formalism. We assume the electric field of incident light $\bm E(t)=E_0 \hat{e}$, where $ \hat{e}=(\hat{i} \cos \theta_p \cos \Omega t + \hat{j} \sin \theta_p \sin \Omega t)$ is the unit polarization vector in which $\theta_p=0 (\pm \pi/4)$ denotes the linearly (circularly) polarized light.

Depending on the nature of light polarization, we can decompose the bulk photovoltaic effect (BPVE) into the piezoelectric-like response (Linearly-polarized (LP) photocurrent), and gyrotropic response (Circularly-polarized (CP) photocurrent) \cite{Sturman_2020,orenstein2021topology}:
\begin{equation} \label{LCP}
	\begin{split}
	J^{(2)}_l&(\omega=0)=\\ &|E_0|^2\sum_{i,j}\int \dfrac{d\Omega}{2\pi} (\beta_{l,ij}^L(\Omega) \text{Re}[e_i e^*_j]+\beta_{lr}^C(\Omega)\kappa_r).
	\end{split}
\end{equation}
The superscripts L and C represent the linear and circular BPVE, respectively. The symmetry of $\beta_{l,ij}^L(\Omega)$ is the same as in the piezoelectric tensor, i.e. $\beta_{l,ij}^L=\beta_{l,ji}^L$, and its form is symmetric under index permutation, then we can use $\beta_{l,ij}^L=\frac{1}{2}[\sigma ^{l;ij}+\sigma ^{l; ji}]$. On the other hand, the vector $\bm \kappa$ is defined as $\bm \kappa=i \bm e \times \bm e^*$ which is non-zero only for circularly polarized light and $\beta_{lr}^C( \Omega)=i \epsilon_{ijr} \sigma^{l;ij}$ is antisymmetric under index permutation and is called the gyration tensor. The \textit{nonlinear Drude response } $\bm{J}^{\text{Dr}}$,\textit{ the Berry curvature dipole} term $\bm{J}^{\text{BCD}}$ , the \textit{injection} stream $ \bm{J}^{\text{Inj}}$, the \textit{ shift} current $\bm{J}^{\text{Sh}}$ and the \textit{gyration} current $\bm{ J}^{\text{Gyr}}$ are classified into the second-order photocurrent response in the DC limit \cite{PhysRevX.11.011001}. With the exception of the Drude response, the other terms are directly governed by various intrinsic geometric quantities of the band structure. Quantum geometries such as Berry curvature and the orbital magnetic moment play key roles in linear electronic and optical transport effects \cite{RevModPhys.82.1539,PhysRevLett.109.181602,RevModPhys.82.1959}. In the nonlinear response,  light-induced direct current or the generation of second harmonics, the other quantum geometries appear in the formalism \cite{PhysRevX.10.041041,bhalla2021quantum,PhysRevLett.112.166601,PhysRevLett.115.216806}. We note that the \textit{metric connection} ($\Gamma$) and the \textit{symplectic connection} ($\tilde{\Gamma}$) have contributions to the shift and gyration currents, and also the quantum geometric \textit{ berry curvature} (${\cal B}$) and \textit{ quantum metric} (${\cal G}$) appear in LP and CP injection terms, respectively.
Furthermore, the berry curvature dipole (BCD) vanishes in an untilted parity violated magnetic Weyl semimetals in which both T and PT symmetries are broken \cite{PhysRevLett.115.216806}. However, it has been shown that a tilt parameter can induce BCD in a T symmetric (non-magnetic) but non-centrosymmetric Weyl systems leading to the accordingly a nonlinear anomalous Hall effect \cite{PhysRevB.103.245119,PhysRevB.97.041101}. The BCD vanishes for an untilted and isotropic Weyl semimetal. In the following sections, we will discuss that the nonlinear interband photocurrent increases near the Weyl crossing points, which is directly attributed to the divergence behavior of quantum geometries near the Weyl nodes, and we will also consider the effect of magnetic dynamics into the photocurrent response.
\section{Theoretical Framework: Quantum kinetic Equation} \label{sec5}
	The quantum Liouville equation of density matrix in the Bloch representation, i.e., $\ket{n,\bm{k}}=\exp(i \bm{k}\cdot \bm{r}) \ket{u_{n,k}}$, is given by
\begin{equation}
	\dfrac{\partial \braket{\rho{(\bm{k},t)}}}{\partial t}+\dfrac{i}{\hbar} [H,\braket{\rho{(\bm{k},t)}}]+\kappa(\braket{\rho{(\bm{k},t)}})=0.
\end{equation}
Here, $H=H_0+H_E$ is the complete Hamiltonian, $\bm{k}$ the crystal momentum, $n$ the band index, $\kappa(\braket{\rho{(\bm{k},t)} })$ is that scatter integral and $H_0$ is the non-interacting Hamiltonian. In the presence of a light-matter interaction, the effect of the electromagnetic field on the time-dependent perturbation can be mapped according to the electric dipole approximation Hamiltonian in the linear approach, i.e. $H_E=e \bm{r} \cdot \bm{E}(t) $, where the electric field couples to the Hamiltonian via a dipole energy. The matrix representation of position operator in the quantum framework would be 
\begin{equation}
	[r_k]_{nm}=i \delta_{nm} \partial_k+{\cal R}_{nm}(k),
\end{equation}
where $\bm{{\cal R}}(k)=\sum\limits_{a=x,y,z} {\cal R}^a(k) \bm{e}_k$ is the Berry vector potential with components $[{\cal R}^a(k)]^{nm}=i\braket{u_k^n|\partial _{k_a} u_k^m}$. This term leads to the topologically non-trivial transport phenomena such as the well-known Hall conductivity in systems with broken TR symmetry. The velocity operator in the Heisenberg picture can be obtained as 
\begin{equation} \label{velocity}
	[v_k]_{nm}=[\dot{r}_k]_{nm}=\hbar^{-1}(\partial_k \epsilon_{n}) \delta_{nm}+i\hbar^{-1} \epsilon_{nm} {\cal R}_{nm}(k),
\end{equation}
where $\epsilon_{nm}=\epsilon_n-\epsilon_m$ is the difference of the eigenvalues of the unperturbed Hamiltonian $H_0$. The second term contributes to the off-diagonal components or the interband responses. If the magnitude of the electric field $|E|$ is small enough, it can be viewed as a perturbation of the Bloch-Hamiltonian operator. Then we may expand the density matrix $\rho$ in powers of the electric field $\rho=\sum\limits_N \rho^{(N)}$, where $\rho^{(N)}$ is the $N^{ th}$ correction to $\rho^{(0)}$ due to the electric field. Then, the quantum kinetic equation in its recursive form would be
\begin{equation} \label{QKE}
	\dfrac{\partial \rho^{(N)}_{(t)}}{\partial t}+\dfrac{i}{\hbar} [H_0,\rho^{(N)}_{(t)}]+\kappa_{nm}(\rho^{(N)}_{(t)})=\dfrac{e \bm{E}(t)}{\hbar} \cdot [{\cal D}_{\bf k} \rho_{(t)}^{(N-1)}].
\end{equation}
Here, ${\cal D}_{\bf k} {\cal O}=\frac{D {\cal O}}{D{\bf k}}=\nabla_{\bf k} {\cal O}-i[\bm{{\cal R}}_{\bf k},{\cal O}]$ is defined as the covariant derivative.
 The solution of Eq. (\ref{QKE}). i.e., the $N^{th}$ order correction to the density matrix, is given by
\begin{equation} \label{rhoN}
	\rho_{nm}^{(N)}(\omega)= e \int \dfrac{d\Omega}{2\pi} d^\omega_{nm} E^i (\Omega) \ [{\cal D}_k^i \rho^{(N-1)}(\omega-\Omega)]_{nm},
\end{equation}
where $d^\omega_{nm}=(\hbar \omega-\epsilon_{nm}+i\hbar \tau^{-1})^{-1}$, and
for simplicity, we estimate the scattering integral as $\kappa_{nm}(\rho^{(N)})\simeq \dfrac{\braket{\rho}}{\tau}$, and $\tau$ is assumed to be $k$-independent.  
 The zeroth correction is the Fermi-distribution function at zero frequency, $\rho_{nm}^{(0)}=2\pi \delta(\omega) \delta_{nm}\  f(\epsilon_{{\bf k}n})$, with $f(\epsilon_{{\bf k}n})=(1+e^{\beta (\epsilon_{{\bf k}n}-\mu)})^{-1}$. The linear-order correction, yields
\begin{equation} \label{rho1}
	\rho^{(1)}_{nm}(\omega)=2\pi e \ d^\omega_{nm} E^i(\omega) \lbrace \partial_{\bf k} f^{(0)}(\epsilon_{{\bf k}n}) \delta_{nm}+i {\cal R}_{\bf k}^{nm} {\cal F}_{nm} \rbrace,
\end{equation}
where ${\cal F}_{nm}\equiv f(\epsilon_{{\bf k}n})-f(\epsilon_{{\bf k}m})$ is defined as the difference between the occupation in bands $n$ and $m$ in equilibrium. The first term in Eq. (\ref{rho1}) is diagonal and captures the intra-band Drude conductivity in metals or doped semimetals with finite Fermi surface, and the second term is off-diagonal and obtains the inter-band optical transitions which in the case of 3D Dirac or Weyl semimetals is linear in $\omega$. 

Using the linear response in Eq. (\ref{rho1}), the second correction $\rho^{(2)}$ would be
\begin{equation} 
	\begin{split}
		\rho_{nm}^{(2)}(\omega)&= e \int \dfrac{d\Omega}{2\pi} d^\omega_{nm} E^i (\Omega) \ [{\cal D}_{\bf k}^i \rho^{(1)}(\omega-\Omega)]_{nm}\\ & =\rho_{\text{I},nm}^{(2)}+\rho_{\text{II},nm}^{(2)}+\rho_{\text{III},nm}^{(2)}.
	\end{split}
\end{equation}
We find $\rho_{nm}^{(1)}(\omega-\Omega)=e \ d^{\omega-\Omega}_{nm} E^i(\omega-\Omega) [{\cal D}_{\bf k}^i \rho^{(0)}]_{nm}$, where we have used Eq. (\ref{rhoN}) and $\rho^{(0)}(\omega-\Omega^\prime-\Omega)=\rho^{(0)} \delta(\omega-\Omega^\prime-\Omega)$. Then, we can decompose the diagonal and off-diagonal parts of the second order correction of density matrix
\begin{equation} \label{rho-21}
	\begin{split}
	&[\rho_{\text{I}}^{(2)}(\omega)]_{nm}=\\
	& e^2 \int \dfrac{d\Omega}{2\pi} d^\omega_{nm} d^{\omega-\Omega}_{nm} E^i(\Omega) E^j(\omega-\Omega) \partial_{k_i} \partial_{k_j} f(\epsilon_{k_n}) \delta_{nm},
		\end{split}
\end{equation}
\begin{equation} \label{rho-22}
	\begin{split}
		& [\rho_{\text{II}}^{(2)}(\omega)]_{nm}=\\
		&i e^2 \int \dfrac{d\Omega}{2\pi} d^\omega_{nm} E^i(\Omega) E^j(\omega-\Omega) \lbrace \partial_{k_i} (d_{nm}^{\omega-\Omega} {\cal R}_{nm}^j {\cal F}_{nm})+ \\
		&i \sum_{n^\prime} ({\cal R}_{k,n^\prime n}^i {\cal R}_{k,n^\prime m}^j d_{n^\prime m}^{\omega-\Omega} {\cal F}_{n^\prime m}-{\cal R}_{k,n n^\prime}^j {\cal R}_{k,n^\prime m}^i d_{n n^\prime}^{\omega-\Omega} {\cal F}_{nn^\prime}) \rbrace,
	\end{split}
\end{equation}
\begin{equation} \label{rho3}
	\begin{split}
	&[\rho_{\text{III}}^{(2)}(\omega)]_{nm}=\\ & i e^2 \int \dfrac{d\Omega}{2\pi} d^\omega_{nm} d^{\omega-\Omega}_{nn} E^i(\Omega) E^j(\omega-\Omega) {\cal R}^i_{k,nm} \partial_{k_j} {\cal F}_{nm}.
	\end{split}
\end{equation}
The summation over the repeating indices $i,j=x,y,z$ is implicit. The terms $\rho_{\text{I}}^{(2)}$ and $\rho_{\text{III}}^{(2)}$ are determined by derivative of the Fermi distribution function, then it would be non-zero for materials with a finite Fermi surface and different velocities in bands $m$ and $n$, e.g. metals or doped semi-metals. The other term $\rho_{\text{II}}^{(2)}$ is also finite without a Fermi surface, so it can exist for insulators as well as for metals or doped semimetals. The N$^\text{th}$ order correction to the light-induced current is obtained by 
\begin{equation}
	\bm{J}^{(N)}_i(\omega)=-e \sum_{n,m} \int \dfrac{d\bm{k}}{(2\pi)^d} v^i_{{\bf k},nm} \ \rho^{(N)}_{{\bf k},mn} (\omega).
\end{equation}
The second-order response can be obtained by setting $N=2$, and using the second-order correction to the density matrix, i.e., $\rho^{(2)}_{{\bf k},mn}$. 
	\begin{figure}[t]
		\centering
		\includegraphics[scale=1.6]{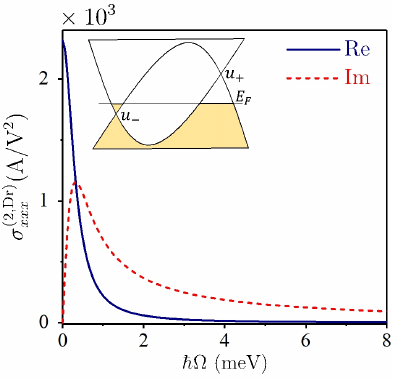} 
		\caption{Non-linear Drude response of inversion asymmetric magnetic Weyl semimetals as a function of light $\Omega$. The real and imaginary parts of the conductivity are demonstrated by blue and red lines, respectively. We set $\tau=1$ ps ($\hbar/ \tau=0.65$ meV), T=10 K ($k_B$T=0.86 meV), $\hbar v_{\rm F}=3.25 \times 10^3$ meV\AA, $\hbar v_F k^w_{x(y)}=140$ meV, $\hbar v_F k^w_{z}=210$ meV, ${\cal J} S=100$ meV, $u_+=50$ meV, $u_-=20$ meV. } \label{fig4}
	\end{figure}
	\section{Intra-band contribution: Non-linear Drude response} \label{sec6}
	The intra-band transitions ($n=m$) in the non-linear response, similar to the linear response, can be only captured for a single band with finite Fermi surface. The term $\rho_{\text{I},nn}^{(2)}$ in Eq. (\ref{rho-21}) is responsible for this non-linear intra-band response.
	Therefore, the non-linear Drude conductivity tensor is given by
	\begin{equation}		\sigma_{l,ij}^{(2,\text{Dr})}=-e^3 \sum_n \int \dfrac{d\bm{k}}{(2\pi)^d} d^0_{nn} d^{-\Omega}_{nn} v_{{\bf k},nn}^l  \partial_{k_i} \partial_{k_j} f(\epsilon_{{\bf k}n}).
	\end{equation}
	We note that the above conductivity is symmetric under $i\leftrightarrow j$, therefore the non-linear Drude conductivity is classified as the LP-photocurrent response. The product of $d^0_{nn} d^{-\Omega}_{nn}$ is a complex function, so the above conductivity tensor can be divided into the real and imaginary parts as
	\begin{equation}
		\begin{split}
			&\text{Re}[\sigma_{l,ij}^{(2,\text{Dr})}]=-\dfrac{e^3}{\hbar}\dfrac{\tau^2}{1+\Omega^2 \tau^2} \ \text{Tr}[ U^{lij}_{nn} f(\epsilon_{{\bf k}n})],\\
			& \text{Im}[\sigma_{l,ij}^{(2,\text{Dr})}]=-\dfrac{e^3}{\hbar} \dfrac{\Omega \tau^3}{1+\Omega^2 \tau^2} \ \text{Tr}[ U^{lij}_{nn} f(\epsilon_{{\bf k}n})],
		\end{split}
	\end{equation}
	where $U^{lij}_{nn}=\braket{n|(\partial_{k_l}\partial_{k_i}\partial_{k_j}H_0(k))|n}$, and $\text{Tr}[{\cal O}]=\sum_{\bm k,n} {\cal O}$. The ratio of the real part to the imaginary part is given by $\dfrac{\text{Re}[\sigma_{l,ij}^{(2,\text{Dr})}]}{\text{Im}[\sigma_{l,ij}^{(2,\text{Dr})}]}=\dfrac{1}{\Omega \tau}$, so the current survives in the limit of $\Omega \tau \ll 1$.
	Figure \ref{fig4} shows the numerical result for the nonlinear Drude response of an inversion-asymmetric magnetic Weyl semimetal with respect to the light frequency $\Omega$. As the figure shows, the nonlinear Drude current is suppressed by dissipation when the light period $1/\Omega$ is shorter than the quasiparticle lifetime $\tau$. Only the intraband response is determined by derivatives of the band energy or the group velocity and does not depend on the quantum geometries. Also, the intraband current does not generate a chiral current, since both nodes are activated at the same time. Therefore, the Drude current cannot rotate the magnetization. In the following, we will discuss that the inter-band contribution to the chiral photocurrent can trigger the magnetization dynamics, resulting in a remarkable impact on the in-plane photocurrent response.
	\section{Inter-band Response} \label{sec7}
The contribution of $\rho_{\text{II},mn}^{(2)}$ (Eq. (\ref{rho-22})) in the photocurrent response results in 
\begin{equation}
	\bm{J}^{(2,\text{II})}_l(\omega=0)= \dfrac{-e}{\hbar} \sum_{n,m} \int \dfrac{d\bm{k}}{(2\pi)^d}  v_{{\bf k},nm}^l  \ \rho_{\text{II},mn}^{(2)}.
\end{equation}
We can decompose Eq. (\ref{rho-22}) into two distinct terms; (1): the first term and the second term with $n^\prime=n,m$ which we denote it by $\rho_{\text{II},1}^{(2)}$ and (2): The second term with $n^\prime \neq n,m$, which we denote it by $\rho_{\text{II},2}^{(2)}$. Then, we have
\begin{equation}
	\begin{split}
&	\rho_{\text{II},1}^{(2)}(\omega=0)\mid_{n,m}=\\ &i e^2 \int \dfrac{d\Omega}{2\pi} d^0_{nm} E^i(\Omega) E^j(-\Omega){\cal D}_{nm}^i (d_{nm}^{-\Omega} {\cal R}_{nm}^j {\cal F}_{nm}),
	\end{split}
\end{equation}
and 
\begin{equation}
	\begin{split}
	&\rho_{\text{II},2}^{(2)}(\omega=0)\mid_{n,m}=-e^2 \sum_{n^\prime \neq n,m} \int \dfrac{d\Omega}{2\pi} d^0_{nm} E^i(\Omega) E^j(-\Omega) \\ &({\cal R}_{{\bf k},n^\prime n}^i {\cal R}_{{\bf k},n^\prime m}^j d_{n^\prime m}^{-\Omega} {\cal F}_{n^\prime m}-{\cal R}_{{\bf k},n n^\prime}^j {\cal R}_{{\bf k},n^\prime m}^i d_{n n^\prime}^{-\Omega} {\cal F}_{nn^\prime}),
	\end{split}
\end{equation}
where ${\cal D}_{nm}^i=\partial_{k_i}-i({\cal R}^i_{nn}-{\cal R}^i_{mm})$ is the U(1)-covariant derivative, and ${\cal F}_{nm}=f(\epsilon_{{\bf k}n})-f(\epsilon_{{\bf k}m})$ is again the difference in distribution function of state $n$ and $m$. Worth noting that the diagonal part of $[\rho_{\text{II},1}^{(2)}]_{nm}$ with $n=m$ vanishes due to ${\cal F}_{nn}=0$, while $[\rho_{\text{II},2}^{(2)}]_{nm}$ exists for both diagonal ($n=m$) and off-diagonal ($n \neq m$) parts. The following subsections include the resultant photocurrent responses given by second-order quantum kinetic formalism.
	\begin{figure}[t]
	\centering
	\includegraphics[scale=0.65]{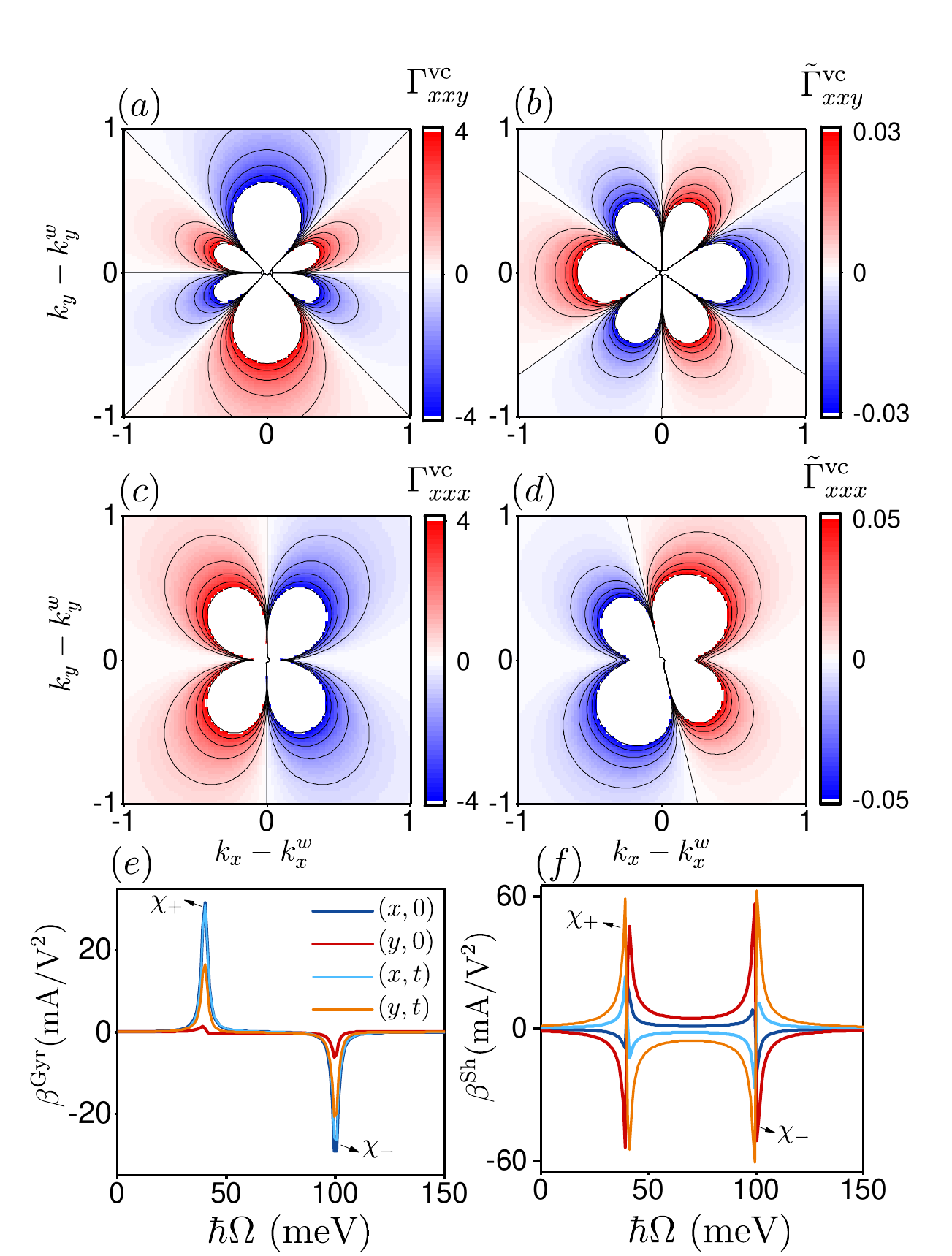} 
	\caption{The momentum-space distribution of quantum geometries for $xxy$ and $xxx$ components of (a),(c): $\Gamma_{vc}$(metric connection) and (b),(d)  $\tilde{\Gamma}^{xxy}_{vc}$(symplectic connection), near the Weyl point with $\chi=+1$. The subscript $v$ and $c$ denote valence and conduction bands, respectively. (e): The frequency dependence of circular gyration response and (f): the linear shift current along the $x$- and $y$- direction where resonance peaks stem from node $\chi=+1$ or $\chi=-1$. $(x(y),0)$: dark blue and red, while $(x(y),t)$: light blue and orange, denote the $x$ or $y$ components of photo-response before and during the magnetic dynamics.} We set $\tau \rightarrow \infty$, $T=10$ K ($k_B$T=0.86 meV), $\hbar v_{\rm F}=3.25 \times 10^3$ meV\AA, ${\cal J} S=500$ meV, $u_+=50$ meV, $u_-=20$ meV. \label{fig5}
\end{figure}
\subsection{Shift current and Gyration Current}
The off-diagonal part of the velocity (second term in Eq. (\ref{velocity})) together with $\rho_{\text{II},1}^{(2)}$ and  $\rho_{\text{II},2}^{(2)}$, lead to the LP shift current, $\beta_{l,ij}^{L,\text{Sh}}$, which originates from the inversion-assymmetric transition of electron position and is definded when $\tau \rightarrow \infty$ \cite{Sturman_2020,PhysRevLett.112.166601}, and CP gyration current response ,$\beta_{lr}^{C,\text{Gyr}}$, satisfying the following expression
\begin{equation}
\begin{split}
&\bm{J}^{(2,\text{II})}_l=\\ & |E_0|^2 \sum_{i,j} \int \dfrac{d\Omega}{2\pi \hbar} \lbrace \beta_{l,ij}^{L,\text{Sh}}(\Omega) e_i e_j^*+i\beta_{lr}^{C,\text{Gyr}}(\Omega) (\bm e \times \bm e^*)_r \rbrace.
\end{split}
\end{equation}
where
\begin{equation}
	\begin{split}
	&\beta_{l,ij}^{L,\text{Sh}}	(\Omega)= \\ &\dfrac{-e^3}{\hbar} \text{Tr}[ \Gamma_{nm}^{lij}  \text{P} \dfrac{1}{\hbar \Omega-\epsilon_{nm}}{\cal F}_{nm}+\pi \tilde{\Gamma}_{nm}^{lij} \delta(\hbar \Omega-\epsilon_{nm}) {\cal F}_{nm}], 
		 	\end{split}
	 \end{equation}
 and
 \begin{equation}  \label{Gyr}
 	\begin{split}
	&\beta_{lr}^{C,\text{Gyr}}(\Omega)= \\ &	-\epsilon_{ijr} \dfrac{e^3}{2\hbar} \text{Tr}[\tilde{\Gamma}_{nm}^{lij} \text{P} \dfrac{1}{\hbar \Omega-\epsilon_{nm}}{\cal F}_{nm}- \pi \Gamma_{nm}^{lij}  \delta(\hbar \Omega-\epsilon_{nm}) {\cal F}_{nm}].
	\end{split}
\end{equation}
 Here, the third-rank metric connection $\Gamma_{nm}^{jli}$, and symplectic connection $\tilde{\Gamma}_{nm}^{jli}$ control the shift and gyration currents near the gap closing point, which are defined as
\begin{equation}
	\begin{split}
		& \Gamma_{nm}^{lij}=\text{Re}[[{\cal D}^l {\cal R}^i_{k}]_{nm} {\cal R}^j_{k,mn}],\\
		& \tilde{\Gamma}_{nm}^{lij}=\text{Im}[[{\cal D}^l {\cal R}^i_{k}]_{nm} {\cal R}^j_{k,mn}]. \\
	\end{split}
\end{equation}
 The integration over 3D $k$-space can be decomposed into an integration over energy and an integration over 2D surface with fixed energy $\epsilon$, i.e., $\int \dfrac{ d^3k}{(2\pi)^3}=\int_0^{\infty}\dfrac{d \epsilon}{2\pi \hbar}\int_\epsilon \dfrac{d^2\sigma}{(2\pi)^2} \dfrac{1}{|v_k|}$, where $v_k$ is the group velocity. The significant contribution arises when the 2D surface in the $k$-space surrounds a Weyl node.
Worth noting that the first term in Eq. \ref{Gyr} could manifest a Fermi-surface effect after using the band-resolved Berry curvature and conducting a partial derivative, $\sum_m \tilde{\Gamma}^{lij}_{nm} F_{nm}=i \sum_m \partial_l {\cal B}^{ij}_{nm} F_{nm}=i {\cal B}^l_n (\partial_l F_{nm})$, where the Berry curvature for the n$^{\text{th}}$ band is defined as 
	$
	{\cal B}^{l}_{n}=\sum_{n^\prime}\dfrac{\epsilon_{lij}}{2} {\cal B}^{ij}_{n n^\prime}=\dfrac{i}{2} \sum_{n^\prime} \epsilon_{lij} [{\cal R}^i_{nn^\prime} {\cal R}^j_{n^\prime n}-{\cal R}^j_{nn^\prime} {\cal R}^i_{n^\prime n}].
	$

 Figure \ref{fig5} demonstrates the different behavior of $xxx$ and $xxy$ components of the metric and symplectic connection near the Weyl nodes. Such nonlinear topological photoresponses arise from the divergence in the quantum geometries near the Weyl points. Therefore, only optical excitations with frequency windows facing the Weyl crossing points make a maximum contribution to the nonlinear photoresponse. Since nodes with opposite chirality have different energies in inversion-asymmetric Weyl semimetals, i.e. $u_+$ and $u_-$, for light frequencies $2\min(u_+,u_-) \leqslant \Omega < 2 \max(u_+, u_-)$ and $u_+\neq -u_-$ only one Weyl node contributes to the interband excitations. The other node plays a role for $\Omega \geqslant 2\max(u_+,u_-)$. Therefore, the optical excitations with different energies are activated at different frequencies when the chemical potential is tuned so that it is not equidistant from two nodes, ie $2\mu \neq u_++u_-$. Figures \ref{fig5} (e) and (f) demonstrate the linear shift and circular gyration currents in the absence and presence of magnetic dynamics. The dynamics of magnetic textures leading to a remarkable change in magnitude and sign of both the $x$ and $y$ components of photocurrents obtained from quantum kinetic theory, corresponding to an in-plane reaction orientation change ( according to Fig. \ref{fig2}(b)). 
	\subsection{Injection Current}
	The diagonal part of the velocity $v_{k,nn}$, together with $\rho_{\text{II},2}^{(2)}$, lead to the injection current in terms of the LP and CP photocurrent 
	\begin{figure}[t]
		\centering
		\includegraphics[scale=0.65]{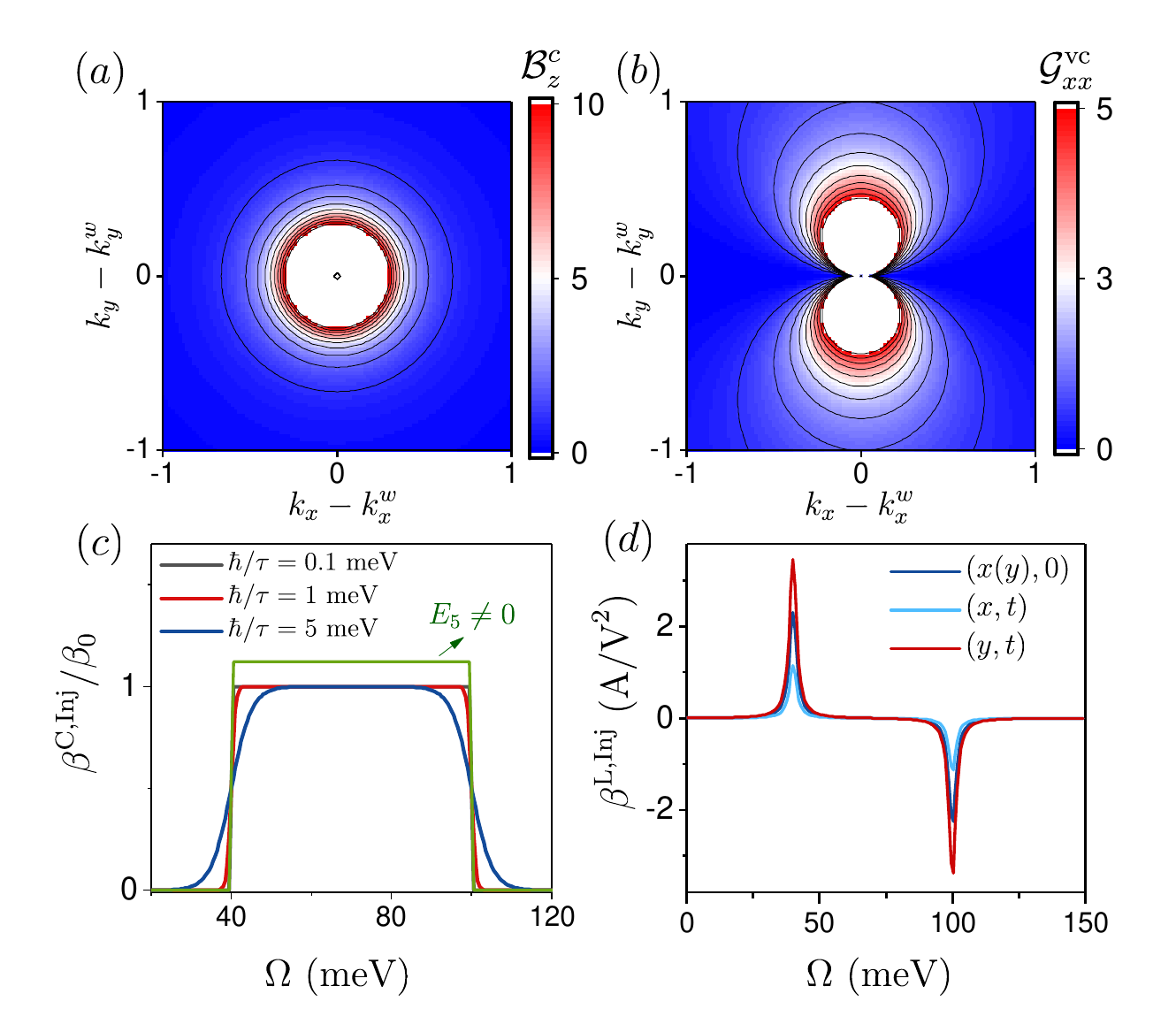} 
		\caption{The momentum-space distribution of quantum geometries (a): ${\cal B}^z_c$ (Berry curvature) and (b): ${\cal G}_{vc}^{xx}$(quantum metric), near the Weyl point with $\chi=+1$. The subscript $v$ and $c$ denote valence and conduction bands, respectively. (c): The quantized circular and (d): the linear injection responses as a function of frequency in the absence and presence of magnetic dynamics (non-zero $\bm E_5$). In (c), the circular injection current
        is a quantized response and vary with the relaxation time. The magnetization dynamics lead to an increase in universal magnitude, which is denoted by a green solid line. In (d), the magnitude of $x$- and $y$- components change, resulting in the rotation of photocurrent. We set $\tau=1$ ps ($\hbar/ \tau=0.65$ meV), T=10 K ($k_B$T=0.86 meV), $\hbar v_{\rm F}=3.25 \times 10^3$ meV\AA, ${\cal J} S=500$ meV, $u_+=50$ meV, $u_-=20$ meV. } \label{fig6}
	\end{figure}
	\begin{equation}
		\bm{J}^{(2,\text{Inj})}_l=|E_0|^2 \sum_{i,j} \int \dfrac{d\Omega}{2\pi} \lbrace \beta_{l,ij}^{L,\text{Inj}} \ \bm e_i \bm e^*_j+i\beta_{lr}^{C,\text{Inj}} (\bm e \times \bm e^*)_r \rbrace,
	\end{equation}
	where 
	\begin{equation}
		\begin{split}
				& \beta_{l,ij}^{L,\text{Inj}}=-\dfrac{\pi \tau e^3}{\hbar} \text{Tr}[\sum_{n^\prime}\Delta_{nn^\prime}^l {\cal G}^{ij}_{n^\prime n} {\cal F}_{n^\prime n}  \delta(\hbar \Omega-\epsilon_{nn^\prime})], \\
				& \beta_{lr}^{C,\text{Inj}}=\epsilon_{ijr} \dfrac{\pi \tau e^3}{2\hbar} \text{Tr}[\sum_{n^\prime}\Delta_{nn^\prime}^l  {\cal B}^{ij}_{n^\prime n}  {\cal F}_{n^\prime n} \delta(\hbar \Omega-\epsilon_{nn^\prime})],
		\end{split}
	\end{equation}
where ${\cal G}^{ij}_{n^\prime n}$ is called the quantum metric and ${\cal B}^{ij}_{n^\prime n}$ is the Berry curvature which are defined as
\begin{equation}
	\begin{split}
		& {\cal G}^{ij}_{n^\prime n}=\dfrac{1}{2}[{\cal R}^i_{nn^\prime} {\cal R}^j_{n^\prime n}+{\cal R}^j_{nn^\prime} {\cal R}^i_{n^\prime n}],\\
		&{\cal B}^{ij}_{n^\prime n}=i[{\cal R}^i_{nn^\prime} {\cal R}^j_{n^\prime n}-{\cal R}^j_{nn^\prime} {\cal R}^i_{n^\prime n}],
	\end{split}
\end{equation}
	, respectively. The injection current clearly depends on the velocity difference along the current response between two bands, topology of bands as well as the relaxation time~\cite{PhysRevB.105.085403}. Similar to the previous linear shift and circular gyration currents, the geometric singularities near the Weyl closing points significantly contribute to the light-induced injection current. Figures \ref{fig6}(a) and (b) represent the quantum metric ${\cal G}^{xx}_{vc}$ and Berry curvature ${\cal B}^{c}_{z}$ near a Weyl node with $\chi=+1$.
	The circular injection current $\beta_{lr}^{C,\text{Inj}}$ is a quantized and constant response that depends on the fundamental and universal quantities like topological charge of Weyl nodes $\chi$, electric charge $e$ and Planck constant $h$: 
		$\sum_{l=x,y,z} \beta_{lz}^{\text{C,Inj}}=-\dfrac{\pi \tau e^3}{h^2}\chi=-\beta_0 \chi \ (\text{A/V}^2)$
\cite{de_Juan_2017}[Fig. \ref{fig6}(c)]. The magnetization dynamics can change the magnitude of $\beta_{xz}$ and $\beta_{yz}$, resulting in enhancement of universal magnitude for circular injection response for magnetic Weyl semimetals [Green dashed line in Fig. \ref{fig6}(c)].
\begin{figure}[t]
		\centering
		\includegraphics[scale=0.6]{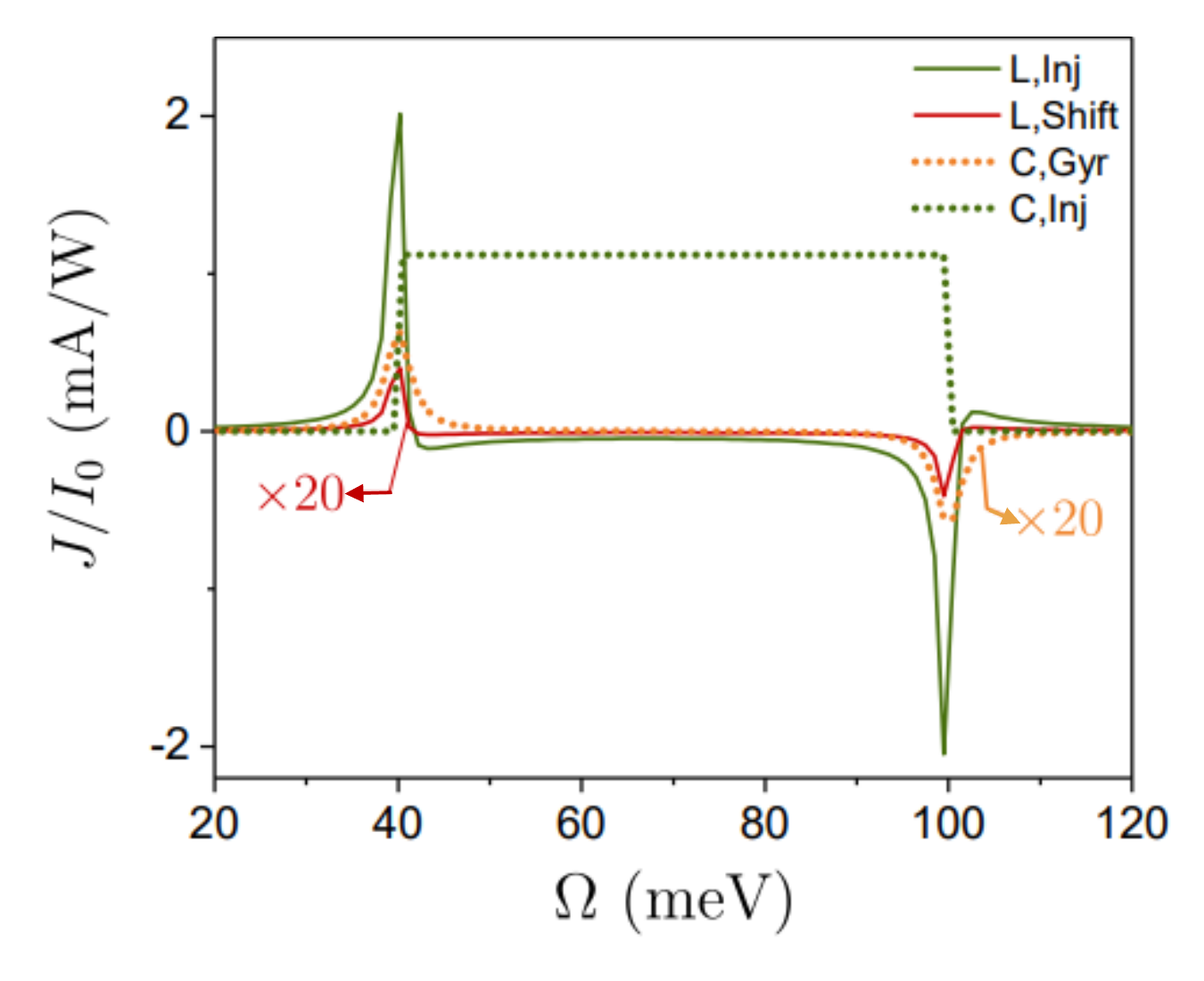} 
		\caption{
		The nonlinear optical response of the linear injection, shift current responses (solid lines), circular gyration and injection currents (dashed line) as a function of light frequency $\Omega$. The linear shift and circular gyration currents (with $\tau\rightarrow \infty$) are multiplied by a factor of 20 due to the graphical purpose. The current $\bm{J}=\sum_{l=x,y,z} J_l$ is the total current where the first and second peaks arise from  node $\chi=+1$ and $\chi=-1$, respectively. $I_0$ is the light intensity in units of W/m$^2$ where we set $|E|^2=2I_0^2/\epsilon_0 c$ and $\epsilon_0 c=\frac{e^2}{4\pi \alpha \hbar}$ with $\alpha=1/137$. Other parameters are the same as Fig.~\ref{fig6}.} \label{fig7}
	\end{figure}
	
Fig. \ref{fig7} collects the results of the nonlinear optical responses of both linearly and circularly polarized lights. Accordingly, the injection currents are much stronger than the shift and gyration currents and then can be the dominant optical response in the system. The photovoltaic current exhibits a strong resonance in order of mA/W in the vicinity of the Weyl nodes, with a magnitude controlled by the momentum relaxation time. Since the Weyl nodes can be arranged by magnetic texture direction, the nonlinear optical response could be used as the basis for a terahertz photo-detector. With some straightforward algebra, we can show that the magnetic dynamics may not be able to notably modify the magnitude of total photocurrent $|\bm{J}|=\sqrt{\sum_l |J_l|^2}$ in Fig. \ref{fig7}, although it results in a change in both direction and magnitude of the in-plane photo-response [Eq. \ref{deltaj}].

Finally, we note that the Pauli blocking is symmetric in an untilted Weyl cone, which means that the $\Omega=2 u_\chi$ frequency window is symmetric. For tilted nodes, the Pauli blocking becomes asymmetric. In other words, the frequency window $\frac{2u_\chi}{1+v_t/v_f}<\Omega<\frac{2u_\chi}{\pm (1-v_t/v_f)}$ becomes wider and more asymmetric for transitions between the bands. The character $\pm$ designates the inclined type I or type II. Furthermore, in a highly asymmetric type II Weyl semimetal a partial compensation between two nodes may occur when the tilt parameter satisfy the condition $\frac{v_t}{v_f}>\frac{1+u_1/u_2}{1-u_1/u_2}$ leading to the partially activation of two nodes simultaneously. The influence of tilting on the nonlinear photoresponse is extensively studied in \cite{PhysRevLett.112.166601,Ma_2017,PhysRevX.10.041041}.
\section{Conclusion} \label{sec8}
This work has investigated the significant correlation of nonlinear DC photocurrent with magnetic texture in a parity-violating magnetic Weyl semimetal. The tunable chemical potential and IS breaking lead to a chiral photocurrent generated by interband transitions of Weyl fermions belonging to a node with chirality $\chi$ while the other node has not yet been activated due to the Pauli blocking.
We have shown that this chiral photocurrent induces STT that causes magnetic texture dynamics, resulting in magnetic texture rotation from the initial $c$-axis to the final $a$- and $b$-axes. The momentum space positions of the Weyl nodes $k_w$ are affected by the magnitude and direction of the magnetization, so any magnetization dynamics can move the Weyl nodes in momentum space, giving them a time dependence of the form $k_w=\bm M (t)$. Accordingly, the dynamic magnetic moments can be mapped to an axial electric field $\bm E_5$. In the dynamic regime of magnetic texture, the presence of an axial electric field induces an additional in-plane current arising from the interplay between non-trivial band topology in momentum space and magnetic texture in real space. Our theory predicts that the in-plane orientation of photocurrents in parity-violating magnetic Weyl semimetals is strongly correlated with the direction of magnetic texture moments. 

	\section{ACKNOWLEDGMENTS}
	 R. A. acknowledges support from the Australian Research Council Centre of Excellence in Future Low-Energy Electronics Technologies (project number CE170100039).
%========================
%\section{ACKNOWLEDGMENT}
%ACKNOWLEDGMENT ACKNOWLEDGMENT ACKNOWLEDGMENT ACKNOWLEDGMENT ACKNOWLEDGMENT ACKNOWLEDGMENT ACKNOWLEDGMENT ACKNOWLEDGMENT ACKNOWLEDGMENT ACKNOWLEDGMENT 
\begin{widetext}
\appendix
\section{Model Hamiltonian and Topological Characteristics} \label{appa}
The low-energy Hamiltonian describing a 3D parity-violating magnetic Weyl semimetal with minimum model with w$_1$ and w$_4$ nodes is given by
\begin{equation} \label{modelH}
		{ H}_0 = \hbar v_{\rm F} \ \tau_z \otimes {\bm \sigma} \cdot ({\bm k}-\bm{k}_\perp^w)
		- \hbar v_{\rm F} \ \tau_0 \otimes (\sigma_z k_z^w- k_M {\bm \sigma} \cdot \hat{\bm M}(r,t))+ u \ \tau_z \otimes \sigma_0 +\lambda \ \tau_y \otimes \sigma_z,
\end{equation}
where $v_{\rm F}$ is the Fermi velocity without tilt, $\bm k^w=(k_\perp^w,k_z^w)$ is the node coordinate in the k-space, the wave-vector $k_M={\cal J}S/\hbar v_{\rm F}$ describes the shift of nodes due to the coupling between electrons and magnetic texture through the exchange interaction ${\cal J}$, and the $2 \times 2$ Pauli matrices $\bm \sigma$ and $\bm \tau$ represents the spin and orbital degree of freedom, respectively. The last two terms are added to the Hamiltonian to violate inversion symmetry (IS), but respect all other symmetries. The term $\lambda \tau_y \otimes \sigma_z$ is the momentum-independent spin-orbit interaction that split the degeneracy at every points except the Weyl crossings. This term is closely analogous to the \textit{Dresselhous spin-orbit interaction} term allowed in the absence of IS. The role of $u\tau_z \otimes \sigma_0$ is shifting two tips of Weyl cones in energy and breaks IS. The inversion operator changes the sign of the momentum and orbital degree of freedom, i.e., $H(k)\rightarrow \tau^x H(k) \tau^x$ and $\tau_x \tau^{y(z)}\tau^x=-\tau^{y,(z)}$. Using the canonical transformation, i.e., $\sigma_{x,y}\rightarrow \tau_z \sigma_{x,y}$, and $\hat{M}\parallel \hat{z}$, the above Hamiltonian is written as
\begin{equation} \label{H0}
		{ H}_{0} =\hbar v_{\rm F} ({\bm k}_\perp-\bm{k}_\perp^w) \ \tau_0 \otimes {\bm \sigma}_\perp
		+\hbar v_{\rm F} \ (k_z \tau_z- (k^w_z-k_M)\tau_0) \otimes \sigma_z +u \ \tau_z \otimes \sigma_0 +\lambda \ \tau_y \otimes \sigma_z,
\end{equation}
In this representation the above Hamiltonian can be presented as a block diagonal Hamiltonian given by $H_0^{\chi}(k)=\hbar v_f (k_\perp-k_\perp^w)\cdot \sigma_\perp+\hbar v_f (\chi k_z-(k_z^w-k_m)) \sigma_z+\chi u_\chi \sigma_0=f_x \sigma_x+f_y \sigma_y+f_{\chi,z} \sigma_z+\chi u_\chi \sigma_0$, where $u_\chi$ determines the shift of nodes in energy due to the IS breaking.
Then, the corresponding eigenvalues are given by $\epsilon_{tk}^\chi=\chi u_\chi +t \sqrt{f_x^2+f_y^2+f^{2}_{\chi, z}}$, where $f_x=\hbar v_{\rm F}(k_x-k_x^w)$, $f_y=\hbar v_{\rm F}(k_y-k_y^w)$, $f_{\chi,z}=\hbar v_{\rm F}(k_z-\chi (k_z^w-k_M))$, and $t=\pm$ denotes the conduction and valence bands, respectively, and $\chi=\pm$ represents the chirality.
The corresponding eigenstates of Eq. (\ref{H0}) would be
$
\ket{n_t^\chi}_k=\dfrac{1}{\sqrt{2}} \begin{pmatrix}
	\sqrt{1+f^\chi_z/\epsilon_{tk}^\chi} \\ \\
	t e^{i\varphi_k} \sqrt{1-f^\chi_z/\epsilon_{tk}^\chi}
\end{pmatrix},
$
where $e^{i\varphi_k}=e^{i\varphi}-\frac{k_\perp^w}{k_\perp} e^{i\varphi_w}$ in which $\varphi$ and $\varphi_w$ are the polar angles of vectors $\bm k$ and $\bm k_w$ in the $k_x-k_y$ plane, respectively.
The Berry connection or Berry vector potential in the eigenstates representation is given by $[{\cal R}_{k,a}]^{nn^\prime}=i \braket{n|\partial_{k_a} n^\prime}$, where $a=x,y,z$ and $n,n^\prime=\pm$ denotes the conduction and valence bands, respectively.
The individual components of the Berry connection are given by 
\begin{equation}
	\begin{split}
		[{\cal R}_{k,x}]^{nn^\prime}=&
		\tilde{\sigma_0} \dfrac{1}{2k_\perp} (\sin \varphi-\frac{k_\perp^w}{k_\perp}\sin \varphi_w)-\tilde{\sigma_z} \dfrac{1}{2k_\perp} \dfrac{f_z^\chi}{\epsilon_k^\chi} (\sin \varphi-\frac{k_\perp^w}{k_\perp} \sin \varphi_w)-\\ & \tilde{\sigma_y} \dfrac{\hbar v_{\rm F} f_z^\chi}{2(\epsilon_k^\chi)^2} (\cos \varphi-\frac{k_\perp^w}{k_\perp} \cos \varphi_w)-\tilde{\sigma_x} \dfrac{\hbar k_{\rm F}}{2\epsilon_k^\chi} (\sin \varphi-\frac{k_\perp^w}{k_\perp} \sin \varphi_w) ,\\
		[{\cal R}_{k,y}]^{nn^\prime}&=-\tilde{\sigma_0} \dfrac{1}{2k_\perp} (\cos \varphi-\frac{k_\perp^w}{k_\perp} \cos \varphi_w)+\tilde{\sigma_z} \dfrac{1}{2k_\perp} \dfrac{f_z^\chi}{\epsilon_k^\chi} (\cos \varphi-\frac{k_\perp^w}{k_\perp} \cos \varphi_w)-\\ & \tilde{\sigma_y} \dfrac{\hbar v_{\rm F} f_z^\chi}{2(\epsilon_k^\chi)^2} (\sin \varphi-\frac{k_\perp^w}{k_\perp} \sin \varphi_w)+\tilde{\sigma_x} \dfrac{\hbar v_{\rm F}}{2\epsilon_k^\chi} (\cos \varphi-\frac{k_\perp^w}{k_\perp} \cos \varphi_w) , \\
		[{\cal R}_{k,z}]^{nn^\prime}&=\tilde{\sigma_y} \dfrac{\hbar v_{\rm F} k_\perp}{2(\epsilon^\chi_k)^2} \dfrac{\partial f_z^\chi (k_z)}{\partial k_z},
	\end{split}
\end{equation}
where $\tilde{\sigma_a}$ are the Pauli matrices in the eigenstates basis to represent the matrix elements of $[{\cal R}_{k,a}]^{nn^\prime}$, i.e., 
$[{\cal R}_{k,a}]^{nn^\prime}=\begin{pmatrix}
	\braket{+|{\cal R}_{k,a}|+} & \braket{+|{\cal R}_{k,a}|-} \\
	\braket{-|{\cal R}_{k,a}|+} & \braket{-|{\cal R}_{k,a}|-}
\end{pmatrix}$.
These components of Berry vectors clearly depend on the specific configuration of Weyl nodes in the Brillouin zone, which is determined by the magnetic-texture properties and lattice structure. The Berry vectors are the building blocks of the topological and geometrical features of the band structure which play the prominent and vital role in photocurrent response. The quantum geometrical quantities also depend on the topological charge or chirality of each node $\chi$.
\begin{figure}[t]
	\centering
	\includegraphics[scale=2.1]{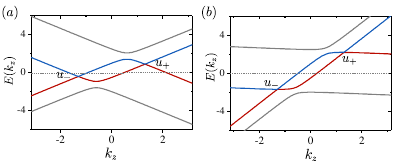} 
	\caption{The band dispersion of model Hamiltonian in Eq. (\ref{modelH}) along the $k_\perp=k_\perp^w$ line as a function of momentum $k_z$ along the Weyl nodes (a) without tilt  $v_t/v_f=0$ and (b): with tilted cones  $v_t/v_f=1$.} \label{fig1p}
\end{figure}
%%%%%%%%%%%%%%%%%%%%%%%%%%%%%%%%%%%%%%%%%%%%%%%%%%%%%%%%%%%%%%%%%%%%%%%%%%%%%%%%
%%%%%%%%%%%%%%%%%%%%%%%%%%%%%%%%%%%%%%%%%%%%%%%%%%%%%%%%%%%%%%%%%%%%%%%%%%%%%%%%
\section{Light-induced magnetic texture dynamics} \label{appb}
In the absence of a magnetic field, the magnetic dynamics based on \textit{Landau-Lifshitz-Gilbert} (LLG) equation \cite{landau1935theory} would be
\begin{equation}
	\begin{split}
	\dfrac{d\hat{M}}{dt}=&\bm{T}_e+\alpha \hat{M} \times \dfrac{d\hat{M}}{dt}=\bm{T}_e+\alpha \hat{M} \times (\bm{T}_e+\alpha \hat{M} \times \dfrac{d\hat{M}}{dt})=\\ & \bm{T}_e+\alpha \hat{M} \times \bm{T}_e+\alpha^2 \hat{M} \times (\hat{M} \times \bm{T}_e)+...=\exp(\alpha \hat{\Theta}) \bm{T}_e
	\end{split}
\end{equation}
where $\hat{\Theta} {\cal O}=\hat{M}\times {\cal O}$ and the spin-transfer torque is given by $\bm{T}_e=\frac{|k_M|}{e \rho_s} \hat{M} \times \bm{J}_5$. For spin-transfer torque to be non-vanishing, the chiral current $J_5$ must be generated in the system. This chiral current can be induced in an inversion asymmetric WSM, since only one Weyl node will be activated due to the Pauli blocking in another node.

The magnetic dynamics can also be understood in terms of an axial electric field $\bm{E}_5=-\frac{\hbar}{e}|k_M| \dot{\hat{M}}=-\frac{\hbar}{e}|k_M| \exp(\alpha \hat{\Theta}) \bm{T}_e$. Therefore, the axial electric field can be written as
\begin{equation} 
	\bm E_5=-\dfrac{\hbar |k_M|^2}{e^2\rho_s} \exp(\alpha \hat{\Theta}) (\hat{M}\times \bm J_5),
\end{equation}

The light-induced time evolution of magnetic texture up to the first order of $\alpha$ is given by
\begin{equation}
	\dfrac{d\hat{M}}{dt}=\bm{T}_e+\alpha \hat{M} \times \bm{T}_e= \dfrac{|k_M|}{e \rho_s} \lbrace \hat{M} \times \bm{J}_5 + \alpha \hat{M} \times (\hat{M} \times \bm{J}_5) \rbrace
\end{equation}
leading to the following equations
\begin{equation}
\begin{split}
	& \dfrac{d\hat{M}_x}{dt}=-\dfrac{|k_M|}{e \rho_s} (J^0_{5,\parallel} \sin \varphi^j_0+\alpha J^0_{5,\parallel} \cos \varphi^j_0), \\
	& \dfrac{d\hat{M}_y}{dt}=\dfrac{|k_M|}{e \rho_s} (J^0_{5,\parallel} \cos \varphi^j_0-\alpha J^0_{5,\parallel} \sin \varphi^j_0), \\
	& \dfrac{d\hat{M}_z}{dt}=0.
\end{split}
\end{equation}
Here, $\varphi^j_0$ is the orientation of in-plane photo-current when $\hat{M}=\hat{z}$. The time evolution of magnetic moments is illustrated by Fig. \ref{fig2}(a).
\section{Inter-band photo-currents} \label{appc}
\subsection{Shift and Gyration Responses} \label{appc1}
	The off-diagonal part of the velocity (second term in Eq. (\ref{velocity})) together with $\rho_{\text{II},1}^{(2)}$ and  $\rho_{\text{II},2}^{(2)}$, leads to the conductivity tensor as following
\begin{equation} \label{Sh0}
	\begin{split}
		\sigma_{l,ij,\text{O}}^{(2,\text{II})}=& \dfrac{e^3}{\hbar} \text{Tr}[d^0_{mn} \epsilon_{nm} {\cal R}^l_{k,nm} {\cal D}_{mn}^i (d_{mn}^{-\Omega}{\cal R}^j_{k,mn}{\cal F}_{nm})]\\ 
		& +i\dfrac{e^3}{\hbar} \text{Tr}[d^0_{mn} \epsilon_{nm} {\cal R}^l_{k,nm} \sum_{n^\prime  \neq m} [f^{ij}_{n^\prime}]_{mn}],
	\end{split}
\end{equation}
where $[f^{ij}_{n^\prime}]_{mn}={\cal R}_{k,n^\prime m}^i {\cal R}_{k,n^\prime n}^j d_{n^\prime n}^{-\Omega} {\cal F}_{n^\prime n}-{\cal R}_{k,m n^\prime}^j {\cal R}_{k,n^\prime n}^i d_{m n^\prime}^{-\Omega} {\cal F}_{mn^\prime}$, and $\text{Tr}=\sum\limits_n \int [d \bm k]$ indicates both a matrix trace and an integration of momentum $\bm k$ over the Brillouin zone, and the summation over $m (\neq n)$ is implicit.
	The term $i{\cal R}^l_{k,nm} \sum\limits_{n^\prime} [f^{ij}_{n^\prime}]_{mn}$ in Eq. (\ref{Sh0}), can be written in a more compact form by using the dummy nature of indices,
	\begin{equation}
		\begin{split}
			i{\cal R}^l_{k,nm} \sum_{n^\prime \neq n \neq m} [f^{ij}_{n^\prime}]_{mn}=&i \sum_{n^\prime \neq n \neq m} {\cal R}^l_{k,nm} {\cal R}_{k,n^\prime m}^i {\cal R}_{k,n^\prime n}^j d_{n^\prime n}^{-\Omega} {\cal F}_{n^\prime n}-i\sum_{n^\prime} {\cal R}^l_{k,nm} {\cal R}_{k,m n^\prime}^j {\cal R}_{k,n^\prime n}^i d_{m n^\prime}^{-\Omega} {\cal F}_{mn^\prime}\\ 
			& =i \sum_{n^\prime \neq n \neq m} ({\cal R}^l_{k,nm}{\cal R}_{k,n^\prime m}^i -{\cal R}_{k,nm}^i {\cal R}^l_{k,mn^\prime}){\cal R}_{k,n^\prime n}^j d_{n^\prime n}^{-\Omega} {\cal F}_{n^\prime n}.
		\end{split}
	\end{equation}
	Using the sum-rule \cite{PhysRevB.52.14636}
	\begin{equation}
		\sum\limits_m i({\cal R}^l_{k,nm}{\cal R}_{k,n^\prime m}^i -{\cal R}_{k,nm}^i {\cal R}^l_{k,mn^\prime})=[{\cal D}^l{\cal R}_k^i]_{nn^\prime}-[{\cal D}^i{\cal R}_k^l]_{nn^\prime},
	\end{equation}
	the second term of Eq. (\ref{Sh0}) is recast as
	\begin{equation} \label{Sh1}
		i\dfrac{e^3}{\hbar} \text{Tr}[d^0_{mn} \epsilon_{nm} {\cal R}^l_{k,nm} \sum_{n^\prime \neq m} [f^{ij}_{n^\prime}]_{mn}]=\dfrac{e^3}{\hbar} \text{Tr}[d^0_{mn} \epsilon_{nm} ([{\cal D}^l{\cal R}_k^i]_{nm}-[{\cal D}^i{\cal R}_k^l]_{nm}) {\cal R}_{k,m n}^j d_{m n}^{-\Omega} {\cal F}_{mn}].
	\end{equation}
	Conducting the partial derivative in the first term of Eq. (\ref{Sh0}) and summing with Eq. (\ref{Sh1}), we find the following expression for $\sigma_{l,ij,\text{O}}^{(2,\text{II})}$
	\begin{equation} \label{condII}
		\sigma_{l,ij,\text{O}}^{(2,\text{II})}=\dfrac{e^3}{2\hbar} \text{Tr}[d^0_{mn} \epsilon_{nm} [{\cal D}^l{\cal R}_k^i]_{nm} {\cal R}^j_{k,mn}  d_{mn}^{-\Omega} {\cal F}_{nm}]+[(i,-\Omega) \leftrightarrow (j,\Omega)].
	\end{equation}
	We introduce the symmetric and anti-symmetric quantities as
	\begin{equation}
		\begin{split}
			& {\cal S}_{nm}^{l;ij}=[ {\cal D}^l {\cal R}^i_{k}]_{nm} {\cal R}^j_{k,mn}+[{\cal D}^l {\cal R}^j_{k}]_{mn} {\cal R}^i_{k,nm}, \\
			& {\cal A}_{nm}^{l;ij}=[ {\cal D}^l {\cal R}^i_{k}]_{nm} {\cal R}^j_{k,mn}-[{\cal D}^l {\cal R}^j_{k}]_{mn} {\cal R}^i_{k,nm}.
		\end{split}
	\end{equation}
	We also define the third-rank geometric tensors such as the \textit{metric connection} $\Gamma_{nm}^{jli}$, and \textit{symplectic connection} $\tilde{\Gamma}_{nm}^{jli}$, which are defined as
	\begin{equation}
		\begin{split}
			& \Gamma_{nm}^{lij}=\text{Re}[[{\cal D}^l {\cal R}^i_{k}]_{nm} {\cal R}^j_{k,mn}],\\
			& \tilde{\Gamma}_{nm}^{lij}=\text{Im}[[{\cal D}^l {\cal R}^i_{k}]_{nm} {\cal R}^j_{k,mn}]. \\
		\end{split}
	\end{equation}
	where ${\cal D}_{nm}^l=\partial_{k_l}-i({\cal R}^l_{nn}-{\cal R}^l_{mm})$. Then we can write ${\cal S}_{nm}^{l;ij}$ and ${\cal A}_{nm}^{l;ij}$ in terms of the geometric quantities
	\begin{equation}
		\begin{split}
			&{\cal S}_{nm}^{l;ij}=\Gamma_{nm}^{lij}+i \tilde{\Gamma}_{nm}^{lij}+(i \leftrightarrow j),\\
			& {\cal A}_{nm}^{l;ij}=\Gamma_{nm}^{lij}+i \tilde{\Gamma}_{nm}^{lij}-(i \leftrightarrow j).
		\end{split}
	\end{equation}
	For the inter-band transitions to dominate, the frequency of light must be larger than the energy difference between the states $n$ and $m$, i.e., $\Omega \geq \epsilon_{nm}\gg 1/\tau$, therefore, the term $\epsilon_{nm} d^0_{mn}d_{mn}^{-\Omega}$ can be decomposed into the real and imaginary parts as
	\begin{equation}
		\epsilon_{nm} d^0_{mn}d_{mn}^{-\Omega}=-\lbrace \text{P} \dfrac{1}{\Omega-\epsilon_{nm}}-i \pi \delta(\Omega-\epsilon_{nm}) \rbrace , \\
	\end{equation}
	where P stands for the principal value in $k$-integration. Then, the conductivity expression in Eq. (\ref{condII}) is simplified as 
	\begin{equation} 
		\sigma_{l,ij,\text{O}}^{(2,\text{II})}=\dfrac{-e^3}{2\hbar} \text{Tr}[{\cal S}_{nm}^{l;ij} \text{P} \dfrac{1}{\Omega-\epsilon_{nm}} {\cal F}_{nm}-i \pi {\cal A}_{nm}^{l;ij} \delta(\hbar \Omega-\epsilon_{nm}) {\cal F}_{nm}].
	\end{equation}
	The above expression can be obtained in terms of LP and CP photocurrent response by using the general formula in Eq. (\ref{LCP})
	\begin{equation}
		\bm{J}^{(2,\text{II})}_l= |E_0|^2 \sum_{i,j} \int \dfrac{d\Omega}{2\pi \hbar} \lbrace \beta_{l,ij}^{L,\text{Sh}}(\Omega) e_i e_j^*+i\beta_{lr}^{C,\text{Gyr}}(\Omega) (\bm e \times \bm e^*)_r \rbrace.
	\end{equation}
	where
	\begin{equation}
		 \begin{split}
				& \beta_{l,ij}^{L,\text{Sh}}(\Omega)=\dfrac{-e^3}{\hbar} \text{Tr}[ \Gamma_{nm}^{lij}  \text{P} \dfrac{1}{\Omega-\epsilon_{nm}}{\cal F}_{nm}+\pi \tilde{\Gamma}_{nm}^{lij} \delta(\hbar \Omega-\epsilon_{nm}) {\cal F}_{nm}], \\
				& \beta_{lr}^{C,\text{Gyr}}(\Omega)=-\epsilon_{ijr} \dfrac{e^3}{2\hbar} \text{Tr}[\tilde{\Gamma}_{nm}^{lij} \text{P} \dfrac{1}{\Omega-\epsilon_{nm}}{\cal F}_{nm}-\pi \Gamma_{nm}^{lij}  \delta(\hbar \Omega-\epsilon_{nm}) {\cal F}_{nm}].
		\end{split}
	\end{equation}
	The LP photocurrent, $\beta_{l,ij}^{L,\text{Sh}}$, is classified into the \textit{Shift current}  \cite{Sturman_2020}, while the CP photo-response, $\beta_{lr}^{C,\text{Gyr}}$, is classified into the \textit{Gyration current}.
	%%%%%%%%%%%%%%%%%%%%%%%%%%%%%%%%%%%%%%%%%%%%%%%%%%%%%%%%%%%%%%%%%%%%
	%%%%%%%%%%%%%%%%%%%%%%%%%%%%%%%%%%%%%%%%%%%%%%%%%%%%%%%%%%%%%%%%%
	\subsection{Injection currents} \label{appc2}
		On the other hand, the diagonal part of the velocity $v_{k,nn}$ together with $\rho_{\text{II},2}^{(2)}$ leads to the following conductivity tensor
	\begin{equation}
		\begin{split}
			\sigma_{l,ij,\text{D}}^{(2,\text{II})}=& \dfrac{e^3}{2\hbar} \text{Tr}[d^0_{mn} v_{k,nm}^l \delta_{nm} \sum_{n^\prime \neq m} [f^{ij}_{n^\prime}]_{mn}]=-i \tau \dfrac{e^3}{2\hbar} \text{Tr}[v_{k,nn}^l \mathop{\sum_{n^\prime}}_{(n^\prime \neq n)} [f^{ij}_{n^\prime}]_{nn}]\\
			& =-i \tau \dfrac{e^3}{2\hbar} \text{Tr}[\sum_{n^\prime} \Delta_{nn^\prime}^l {\cal R}^i_{nn^\prime} {\cal R}^j_{n^\prime n} d^{-\Omega}_{n^\prime n} {\cal F}_{n^\prime n}]+[(i,-\Omega) \leftrightarrow (j,\Omega)]\\
			&=-i \tau \dfrac{e^3}{2\hbar} \text{Tr}[\sum_{n^\prime} \Delta_{nn^\prime}^l {\cal R}^i_{nn^\prime} {\cal R}^j_{n^\prime n} (d^{-\Omega}_{n^\prime n}+d^{\Omega}_{n n^\prime}) {\cal F}_{n^\prime n}]
		\end{split}
	\end{equation}
	where we have used $[f^{ij}_{n^\prime}]_{nn}={\cal R}_{k,n^\prime n}^i {\cal R}_{k,n^\prime n}^j d_{n^\prime n}^{-\Omega} {\cal F}_{n^\prime n}-{\cal R}_{k,n n^\prime}^j {\cal R}_{k,n^\prime n}^i d_{n n^\prime}^{-\Omega} {\cal F}_{nn^\prime}$, and $\Delta_{nn^\prime}^l=v_{n}^l-v_{n^\prime}^l$ is the group velocity difference between the band $n$ and $n^\prime$, $\text{Tr}=\sum\limits_n \int [d \bm k]$. 
	
	The real and imaginary parts of the quantity ${\cal R}^i_{nn^\prime} {\cal R}^j_{n^\prime n}$ can be defined in terms of symmetric and antisymmetric quantum geometric quantities as \cite{bhalla2021quantum}
	\begin{equation}
		{\cal R}^i_{nn^\prime} {\cal R}^j_{n^\prime n}={\cal G}^{ij}_{n^\prime n}-i {\cal B}^{ij}_{n^\prime n}/2,
	\end{equation}
	where ${\cal G}^{ij}_{n^\prime n}$ is called the \textit{quantum metric} and ${\cal B}^{ij}_{n^\prime n}$ is the \textit{Berry curvature} which are defined, respectively
	\begin{equation}
		\begin{split}
			& {\cal G}^{ij}_{n^\prime n}=\dfrac{1}{2}[{\cal R}^i_{nn^\prime} {\cal R}^j_{n^\prime n}+{\cal R}^j_{nn^\prime} {\cal R}^i_{n^\prime n}],\\
			&{\cal B}^{ij}_{n^\prime n}=i[{\cal R}^i_{nn^\prime} {\cal R}^j_{n^\prime n}-{\cal R}^j_{nn^\prime} {\cal R}^i_{n^\prime n}].
		\end{split}
	\end{equation}
	The quantity ${\cal B}^{ij}_{n^\prime n}$ is related to the Berry curvature for the n$^{\text{th}}$ band as
	\begin{equation} \label{BC}
		{\cal B}^{l}_{n}=\sum_{n^\prime}\dfrac{\epsilon_{lij}}{2} {\cal B}^{ij}_{n n^\prime}=\dfrac{i}{2} \sum_{n^\prime} \epsilon_{lij} [{\cal R}^i_{nn^\prime} {\cal R}^j_{n^\prime n}-{\cal R}^j_{nn^\prime} {\cal R}^i_{n^\prime n}].
	\end{equation}
	Using the fact that $d^{-\Omega}_{n^\prime n}+d^{\Omega}_{n n^\prime}=-2i \delta(\hbar \Omega-\epsilon_{nn^\prime})$, The conductivity $\sigma_{l,ij,\text{D}}^{(2,\text{II})}$ is written as
	\begin{equation} \label{inj0}
		\sigma_{l,ij,\text{D}}^{(2,\text{II})}=-\dfrac{\pi \tau e^3}{\hbar} \text{Tr}[\sum_{n^\prime}\Delta_{nn^\prime}^l {\cal F}_{n^\prime n} ( {\cal G}^{ij}_{n^\prime n} -\dfrac{i}{2} {\cal B}^{ij}_{n^\prime n}) \delta(\hbar \Omega-\epsilon_{nn^\prime})]
	\end{equation}
	The term contains ${\cal G}^{ij}_{n^\prime n}$ (${\cal B}^{ij}_{n^\prime n}$) satisfies the symmetric (anti-symmetric) condition under permutation $i\leftrightarrow j$, then it is classified into LP (CP) current, known as the \textit{injection current} arising from the longitudinal velocity injection.
	
	Then, the injection current response in terms of general expression for LP and CP photocurrent would be
	\begin{equation}
		\bm{J}^{(2,\text{Inj})}_l=|E_0|^2 \sum_{i,j} \int \dfrac{d\Omega}{2\pi} \lbrace \beta_{l,ij}^{L,\text{Inj}} \ \bm e_i \bm e^*_j+i\beta_{lr}^{C,\text{Inj}} (\bm e \times \bm e^*)_r \rbrace,
	\end{equation}
	where 
	\begin{equation}
	\begin{split}
				& \beta_{l,ij}^{L,\text{Inj}}=-\dfrac{\pi \tau e^3}{\hbar} \text{Tr}[\sum_{n^\prime}\Delta_{nn^\prime}^l {\cal G}^{ij}_{n^\prime n} {\cal F}_{n^\prime n}  \delta(\hbar \Omega-\epsilon_{nn^\prime})], \\
				& \beta_{lr}^{C,\text{Inj}}=\epsilon_{ijr} \dfrac{\pi \tau e^3}{2\hbar} \text{Tr}[\sum_{n^\prime}\Delta_{nn^\prime}^l  {\cal B}^{ij}_{n^\prime n}  {\cal F}_{n^\prime n} \delta(\hbar \Omega-\epsilon_{nn^\prime})].
		\end{split}
	\end{equation}
\end{widetext}
\bibliography{Ref}
\end{document}